\documentclass[twocolumn]{article}

\usepackage[english]{babel}
\usepackage[a4paper,top=2cm,bottom=2cm,left=1.5cm,right=1.5cm,marginparwidth=1.75cm]{geometry}

\usepackage{amsmath}
\usepackage{graphicx}
\usepackage[colorlinks=true, allcolors=blue]{hyperref}
\usepackage[T1]{fontenc}
\usepackage{graphicx}
\usepackage{graphbox}
\usepackage{algorithm}
\usepackage[noend]{algpseudocode}
\usepackage{amsfonts}
\usepackage{amsmath}
\usepackage{array}
\usepackage{mathtools}
\usepackage{xcolor}
\usepackage[numbers]{natbib}
\usepackage{mathtools}
\usepackage{stmaryrd}
\usepackage{amssymb}
\usepackage{moreverb}
\usepackage{amssymb}
\usepackage{lineno}
\usepackage{graphicx}
\usepackage{amsmath}
\usepackage{moreverb}
\usepackage{amssymb}
\usepackage{lineno}
\usepackage{graphicx}
\graphicspath{ {images/} }
\usepackage{multirow}
\usepackage[utf8]{inputenc}
\usepackage{mathtools}
\usepackage{microtype}
\usepackage{lipsum}
\usepackage{setspace}
\let\oldnl\nl
\newcommand{\nonl}{\renewcommand{\nl}{\let\nl\oldnl}}
\usepackage{stackengine}
\usepackage{float}
\usepackage{stmaryrd}
\usepackage{multirow}
\usepackage{epstopdf}
\usepackage{makecell}
\usepackage{multirow}
\usepackage{graphicx}
\newcommand\keywords[1]{%
	\begingroup
	\let
	\par
	\noindent\textbf{Keywords:}\\#1\par
	\endgroup
}

\newcommand\blfootnote[1]{%
	\begingroup
	\renewcommand\thefootnote{}\footnote{#1}%
	\addtocounter{footnote}{-1}%
	\endgroup
}

\title{\textbf{iFogSim2: An Extended iFogSim Simulator for Mobility, Clustering, and Microservice Management in Edge and Fog Computing Environments}}

\author{Redowan Mahmud, Samodha Pallewatta, Mohammad Goudarzi, and Rajkumar Buyya}

\begin{document}
\maketitle

\begin{abstract}
Internet of Things (IoT) has already proven to be the building block for next-generation Cyber-Physical Systems (CPSs). The considerable amount of data generated by the IoT devices needs latency-sensitive processing, which is not feasible by deploying the respective applications in remote Cloud datacentres. Edge/Fog computing, a promising extension of Cloud at the IoT-proximate network, can meet such requirements for smart CPSs. However, the structural and operational differences of Edge/Fog infrastructure resist employing Cloud-based service regulations directly to these environments. As a result, many research works have been recently conducted, focusing on efficient application and resource management in Edge/Fog computing environments. Scalable Edge/Fog infrastructure is a must to validate these policies, which is also challenging to accommodate in the real-world due to high cost and implementation time. Considering simulation as a key to this constraint, various software has been developed that can imitate the physical behaviour of Edge/Fog computing environments. Nevertheless, the existing simulators often fail to support advanced service management features because of their monolithic architecture, lack of actual dataset, and limited scope for a periodic update. To overcome these issues, we have developed multiple simulation models for service migration, dynamic distributed cluster formation, and microservice orchestration for Edge/Fog computing in this work and integrated with the existing \textit{iFogSim} simulation toolkit for launching it as \textit{iFogSim2}. The performance of iFogSim2 and its built-in policies are evaluated using three use case scenarios and compared with the contemporary simulators and benchmark policies under different settings. Results indicate that the proposed solution outperform others in service management time, network usage, ram consumption, and simulation time.	\blfootnote{\textbf{R. Mahmud} is with The School of Computing Technologies, STEM College, RMIT University, Melbourne, Australia}
\blfootnote{\textbf{S. Pallawetta, M. Goudarzi, and R. Buyya} are with The Cloud Computing and Distributed Systems (CLOUDS) Laboratory, School of Computing and Information Systems, The University of Melbourne, Australia}
\blfootnote{\textbf{Corresponding Author}: Mohammad Goudarzi, Email: mgoudarzi@student.unimelb.edu.au}
\end{abstract}


	\section{Introduction}
The Internet of Things (IoT) paradigm has drastically changed the convention of interactions between physical environments and digital infrastructures that sets the tone of using numerous IoT devices including fitness trackers, voice controllers, smart locks, and air quality monitors in our daily activities. Currently, IoT devices are contributing 11.5 Zetta bytes to the total data generated around the globe, which is experiencing an exponential rise each year \cite{WEF}. The recent adoption of Edge/Fog computing has relaxed the requirements of harnessing Cloud datacentres for different IoT-driven use cases. This novel computing paradigm spans the computing facilities across the proximate network and enables smart health, smart city, Agtech, Industry 4.0, and other IoT-enabled Cyber-Physical Systems (CPSs) to run the necessary application software in the vicinity of the data source \cite{Afrin}. Thus, Edge/Fog computing ensures the delivery of application services to the requesting IoT-enabled CPSs with reduced data transmission and propagation delay and lessens the possibility of congestion within the core network infrastructure by inhibiting the transfer of a large amount of raw data to the Cloud datacentres.  
	\par The realisation of Edge/Fog computing environments primarily depends on the integration of computing resources such as processing cores, nano servers, and micro datacentres with the traditional networking components, including gateway routers, switches, hubs, and base stations \cite{Con-Pi,deng2021fogbus2}. In contrast to Cloud datacentres, such Edge/Fog computing nodes are highly distributed. Similarly, the heterogeneity in terms of resource architecture, communication standards, and operating principles predominantly exist among these nodes. Because of such constraints, the centralised Cloud-based resource provisioning and application placement techniques have compatibility issues with Edge/Fog computing environments and cannot be applied directly to regulate the respective services \cite{satish,goudarzi2019fog}. Identifying this potential research gap, a significant number of initiatives have been taken to develop efficient service management policies for Edge/Fog computing. The research interest for Edge/Fog computing has increased around 69\% in the last five years \cite{trend}. However, the newly developed service management policies for Edge/Fog computing environments require extensive validation before enterprise adoption. 
	\par Real-world deployment is the most effective approach to evaluate the performance of any service management policy. However, since Edge/Fog computing environments incorporate numerous IoT devices and computing nodes, both in tiered and flatted order with vast amounts of batch or streaming data and distributed applications, their real-world implementation with such a scale is challenging. The lack of global Edge/Fog service providers offering infrastructure on pay-as-you-go models like commercial Cloud platforms such as Microsoft Azure and Amazon AWS further forces researchers to set up real Edge/Fog computing environments by themselves for costly policy evaluation. Additionally, the implementation time for a real-world environment is significantly high, and the modification and tuning of any entity or system parameters during the experiments are tedious \cite{Edge-Affinity}. These constraints can be addressed by simulating Edge/Fog computing environments. Not only does simulation provide support for designing a customized and scalable experiment environment, but it also assists in the repeatable evaluation under different settings. 
	\par Although there exist several simulators for Edge/Fog computing environments, a majority of them lack benchmarks to validate other service management policies. They merely use synthetic data without any functional ground, which often direct to biased and erroneous performance evaluation. Their monolithic architecture also refuses periodic updates, resisting them to cope up with the advanced features of genuine Edge/Fog computing nodes. Consequently, they fail in imitating various complex scenarios triggered by uncertain device mobility, resource constraints, and heavy-weight computations. To meet such shortcomings, a set of simulation models for mobility-aware application migration, dynamic distributed cluster formation, and microservice orchestration has been developed in this work. The proposed models exploit real-world dataset and are comprehensively mimics the capabilities of the state-of-the-art Edge/Fog computing nodes and IoT devices. These models are integrated with the existing \textit{iFogSim} simulator \cite{ifogsim} and launched as \textit{iFogSim2} for widespread adoption as benchmarks. The major contributions of our work are listed below. 

	\begin{itemize}
		\item A service migration simulation model that can operate across multi-tier infrastructure and support simplified integration of real-world dataset. The launching version uses EUA Dataset as the default and accommodates different device mobility models, including pathway and random waypoint.   
		\item A dynamic distributed cluster formation among multi-tier infrastructure is proposed, where Edge/Fog nodes in different tiers can provide services with a higher quality of service. The cluster management is performed in a distributed manner through which different cluster formation policies can be simultaneously integrated.   
		\item An orchestration model for microservices deployed across multi-tier infrastructure, which enables placement policies to dynamically scale microservices among federated Edge/Fog nodes to improve resource utilisation. Different service discovery and load balancing policies can be integrated to simulate the dynamic microservice behavior.
	\end{itemize}
	\par The rest of the paper is organized as follows. In Section \ref{sec_related}, related researches are reviewed. Section \ref{sec_extension} denotes how the proposed simulation models are integrated with the iFogSim simulator. The performance of proposed iFogSim2 simulator is evaluated in Section \ref{sec_performance}. Finally, Section \ref{sec_conclusion} concludes the paper with future works.
	\section{Related work} \label{sec_related}
	\par Among the existing simulators for Edge/Fog computing, the \textit{EdgeCloudSim} software supports the nomadic movements of the IoT devices \cite{Sonmez}. Additionally, it considers the static deployment and coverage area for the gateway nodes and assumes the link quality between IoT and gateway nodes remains always the same despite their distance. Similarly, the \textit{FogNetSim++} simulator developed by Qayyum et al. \cite{Qayyum} can imitate different mobility models for IoT devices, including random waypoint, mass mobility, and linear mobility. It also provides the facilities to develop customized mobility models as per the operating environment. The mobility support system of FogNetSim++ is loosely coupled with the core simulation engine, and thus its extension requires the least modifications of the primary libraries. However, both EdgeCloudSim and FogNetSim++ lack abstractions for implementing microservice orchestration and dynamic clusterisation among multiple Edge/Fog nodes. In \cite{Jha}, Jha et al. proposed another simulator named \textit{IoTSim-Edge} for modeling the characteristics of IoT devices in the Fog computing environment. It represents IoT applications as a collection of microservices, and the mobility model associated with IoTSim-Edge incorporates different attributes of IoT devices, including its range, velocity, location, and time interval. This simulator also facilitates users to implement their mobility model by extending the core simulator programming interfaces but barely highlights the clustering of the computing nodes. 
	
	\par Furthermore, Puliafito et al. \cite{Puliafito2} have recently developed \textit{MobFogSim} for simulating device mobility and application migration in Fog computing environments. It is an extension of iFogSim that modifies the basic functionalities of different iFogSim components with mobility features. However, the mobility support system of MobFogSim only deals with the IoT gateways and Cloud datacentres instead of tiered Edge/Fog infrastructure and limits the scope for creating clusters in Edge/Fog computing environments. Mechalikh et al. \cite{Mechalikh} developed another simulator called \textit{PureEdgeSim} to evaluate the performance of Fog and Cloud computing environments for different IoT-driven use cases. The mobility support system of PureEdgeSim includes a location manager, which is loosely coupled with the core simulation engine. However, the default mobility-aware application management policy of PureEdgeSim is complex and difficult to customize. It also has limitations in forming node clusters and augmenting microservice management techniques. Conversely, Mass et al. \cite{Mass} developed the \textit{STEP-ONE} simulator to imitate the operations of Fog-based opportunistic network environments. STEP-ONE extends the conventional ONE simulator with advanced mobility and messaging interfaces and primarily focuses on modeling simple business processes. Although STEP-ONE incorporates support for the real-world dataset, it lacks default policies for mobility management, node clustering, and microservice orchestration. Likewise, in \cite{Lera}, Lera et al. proposed \textit{YAFS} simulator for Fog computing to design and deploy various IoT applications with customized resource management policies. The mobility support system of YAFS operates based on the sender-receiver relationship between the Fog nodes that identifies the shortest path during device movements. YAFS also defines logical relations among microservices through graphs and provides interfaces for node clustering. 
	\begin{table*}[!t]
		\scriptsize 
		\centering
		\caption{A Summary of related work and their comparison}\label{Tab:summary} 
		\begin{tabular}{|p{2.5 cm}|p{1.5 cm}|p{2 cm}|p{2.4 cm}|p{2 cm}|p{1.5 cm}|p{2.4cm}|}
			\hline
			Simulators & Real dataset  & Benchmark policy & \multicolumn{3}{c|}{Supports} & Modular architecture\\
			\cline{4-6}
			& & & Customised mobility & Cluster formation & Microservices & \\\hline
			EdgeCloudSim \cite{Sonmez} &  & \checkmark & &  &  &  \\
			FogNetSim++ \cite{Qayyum} &  &  & \checkmark &  &  & \checkmark \\
			IoTSim-Edge \cite{Jha} &  & \checkmark & \checkmark &  & \checkmark &  \\
			MobFogSim \cite{ Puliafito2} & \checkmark & \checkmark & \checkmark &  & \checkmark &  \\
			PureEdgeSim \cite{Mechalikh} &  & \checkmark &  &  &  & \checkmark \\
			STEP-ONE \cite{Mass} & \checkmark &  & \checkmark &  &  & \\
			YAFS \cite{Lera} &  & \checkmark &  & \checkmark & \checkmark & \checkmark \\
			IoTNetSim \cite{Salama} & \checkmark &  & \checkmark &  &  & \checkmark \\
			SatEdgeSim \cite{Wei}  & & \checkmark & \checkmark &  &  & \checkmark \\
			ECSNeT++ \cite{Amarasinghe} & \checkmark & \checkmark &  &  & \checkmark & \checkmark \\
			IoTSim-Osmosis \cite{Alwasel} & & \checkmark &  &  & \checkmark &  \\
			\textbf{iFogSim2} &  \checkmark &  \checkmark & \checkmark & \checkmark & \checkmark & \checkmark  \\\hline
		\end{tabular}  
	\end{table*}  
	\par Furthermore, the IoTNetSim \cite{Salama} simulator for Edge/Fog computing environments developed by Salama et al. can model different IoT devices and their granular details, including energy profile. It supports the mobility of IoT devices in three-dimensional space. Although IoTNetSim is highly modular, it lacks benchmark policies for mobility-driven service management and dynamic cluster formation. Wei et al. proposed another simulator named SatEdgeSim for evaluating the performance of service management policies in three-tier satellite edge computing environments \cite{Wei}. Considering the high mobility of satellite nodes, It supports the dynamic alteration in network topology and imitates the impact of communication distance on service offloading delay. Although SatEdgeSim is modular, it barely exploits the concept of microservice. Conversely, the IoTSim-Osmosis simulator, developed by Alwasel et al. targets the migration of workload to edge nodes based on performance and security requirements. It considers the IoT environment as a four-tier architecture and models the applications in form of microservices. However, the simulation components of IoTSim-Osmosis are tightly coupled and constrained in imitating device mobility. ECSNeT++ \cite{Amarasinghe} is another simulator developed by Amarasinghe et al. that mimics the execution of distributed stream processing (DSP) applications in Edge/Fog computing environments. It extends OMNeT++/INET and provides multiple configurations for two real DSP applications with calibration and deployment management policy. Nevertheless, ECSNeT++ lacks interfaces for supporting customized mobility of IoT devices and forming dynamic clusters.    
	\begin{figure*}[!t]
		\includegraphics[width=0.9\textwidth]{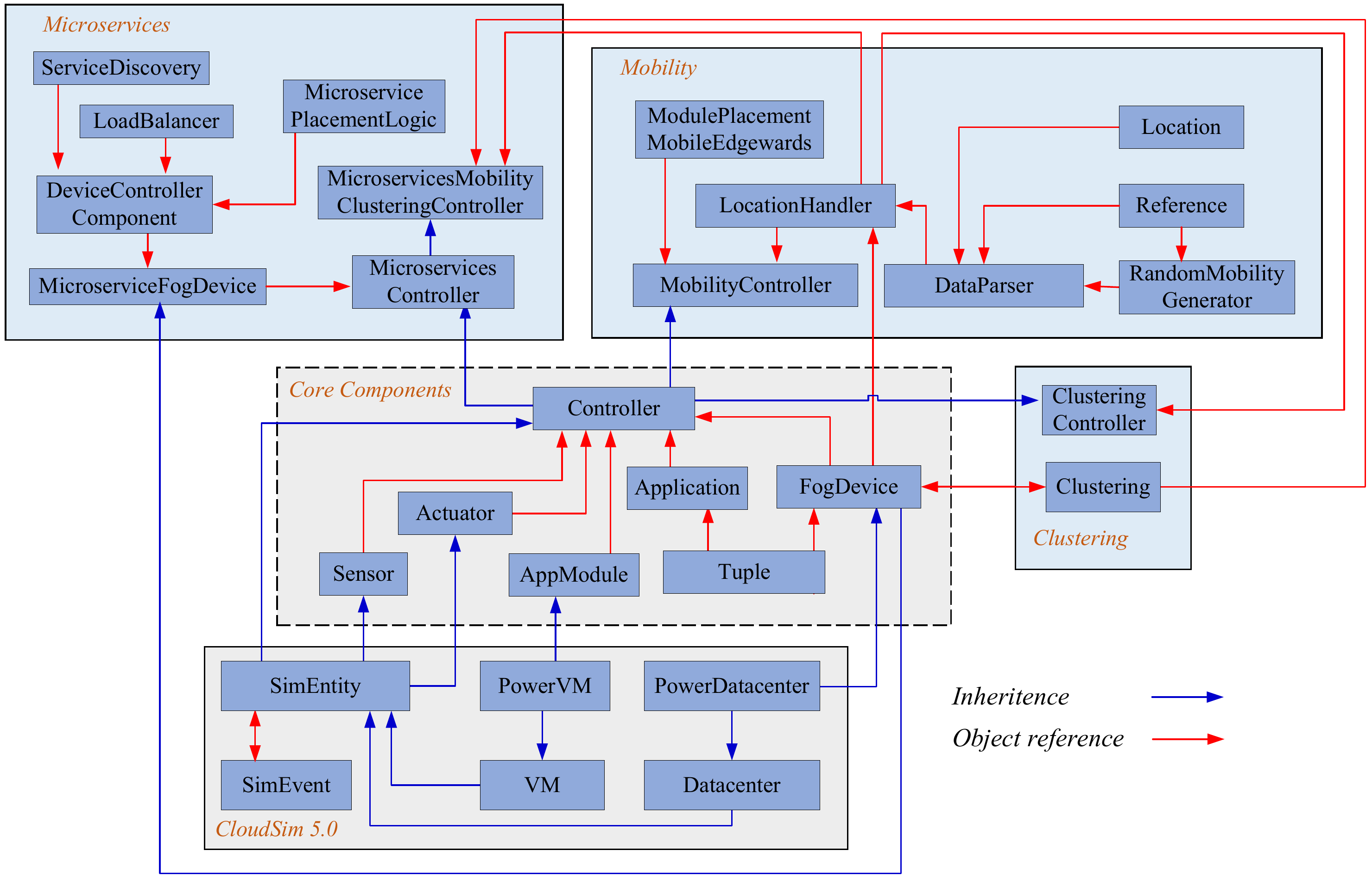}
		\caption{Overview of iFogSim2 simulator}
		\label{Fig:software}
	\end{figure*}
	\par In comparison to most of the existing solutions, the proposed iFogSim2 simulator simultaneously supports the integration of real dataset for evaluating the performance of different service management policies in Edge/Fog computing environments and provides default techniques for mobility management, node clustering, and microservice orchestration, which can be adopted as benchmarks during performance comparison. Additionally, the simulation components of iFogSim2 are highly modular that eases its customisation for imitating a wide range of service management scenarios in Edge/Fog computing environments.        
	\section{iFogSim2 components} \label{sec_extension}
	To address the prevailing limitations of the iFogSim simulator in supporting the migration of application, logical grouping of Fog nodes, and orchestration of loosely-coupled application services, three new components, namely \textit{Mobility}, \textit{Clustering} and \textit{Microservices}  have been implemented and included in iFogSim2 (as shown in Fig. \ref{Fig:software}). 
	
	\par While simulating any use case through basic iFogSim, its \textit{Controller} class contains the object references of all iFogSim core classes such as \textit{FogDevice}, \textit{Sensor}, \textit{Actuator} and \textit{AppModule}. Through an \textit{Application} object, the \textit{Controller} class can also access the \textit{Tuple} class of iFogSim. Therefore, in the newly developed three components, we have inherited the \textit{Controller} class separately, so that they can be easily integrated with the core iFogSim simulator. Additionally, the \textit{Controller} class of iFogSim itself is a subclass of \textit{SimEntity}, which also helps bridge iFogSim2 with the core CloudSim 5.0 simulator. Besides, \textit{FogDevice} is updated with several parameters to support the integration of these new components. In the following subsections, different iFogSim2 components are discussed in detail.
	
	\subsection{Mobility}
	The Mobility of IoT devices can affect the performance of Edge/Fog computing, especially when they change access points very frequently \cite{5G}. This event urges to migrate the requested application service of the IoT devices from one computing node (migration source) to another (migration destination) for ensuring the committed QoS. In a logical multi-tier computing infrastructure like Edge/Fog, such service migration operations depend on the following aspects. 
	\begin{itemize}
		\item The location of IoT devices. 
		\item The timeline of movement or the mobility direction and speed of the device. 
		\item The identification of an intermediate node to which the migration source can upload the corresponding application service and the migration destination can download; provided that there is no direct link between the respective migration points.   
	\end{itemize} 
	Based on them, the performance indicators of the Edge/Fog environment such as network delay, energy consumption, and service delivery time can also vary significantly. Therefore, taking these facts into account and aiming to assist users in customizing them, we have developed several classes in the \textit{Mobility} component of iFogSim2. A detailed description of these classes is given below.
	\par \textit{DataParser:} This class works as the interface for extending data from external sources to the proposed iFogSim2 components. Currently, it incorporates abstractions for reading data from $.csv$ files, which can be further extended for other formats. However, while simulating mobility-driven cases, iFogSim2 adopts the EUA Datasets which contains the location information of a notable number of Edge/Fog nodes deployed across Central Business District (CBD) regions of major cities in Australia, including Melbourne and Sydney. 
	\par
\begin{figure*}[!ht]
		\centering
		\begin{minipage}[t]{0.3\textwidth}
			\centering
			\vspace{0pt}
			\includegraphics[width=\textwidth]{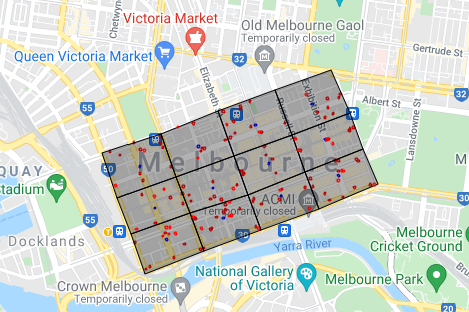}
		\end{minipage}
		\qquad
		\hfill
		\begin{minipage}[t]{0.3\textwidth}
			\centering
			\vspace{0pt}
			\includegraphics[width=\textwidth]{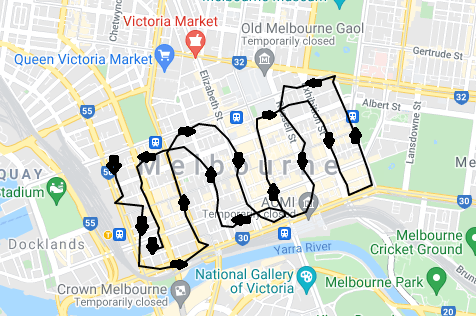}
		\end{minipage}
		\qquad
		\hfill
		\begin{minipage}[t]{0.3\textwidth}
			\centering
			\vspace{0pt}
			\includegraphics[width=\textwidth]{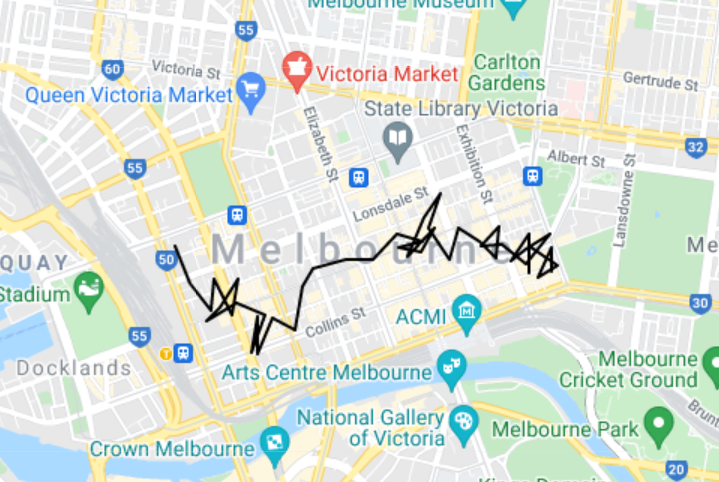}
		\end{minipage}
		\footnotesize{(a)\hspace{160pt}(b)\hspace{160pt}(c)}\\
		\caption{(a) Block-wise Edge/Fog computing nodes, (b) Directional Movement of an user, and (c) Random Movement of an user in Melbourne Central Business District}
		\label{Fig:mobility}
	\end{figure*}

 To ensure granularity, we further customised the dataset by segmenting the respective regions in multiple blocks and selecting a particular node at the middle of each block as the proxy server. All nodes but the proxy server within a block act as the gateway for the IoT devices. As a means of notations, proxy servers are specified as the tier-1 nodes in iFogSim2 and assumed to be the immediate upper tier contact for the gateways residing at the same block. In this case, the gateways are referred to as tier-2 nodes. For example, Fig. \ref{Fig:mobility}.a presents the location of tier-1 (marked in blue) and tier-2 (marked in red) nodes deployed in the Melbourne CBD. Finally, such a logical hierarchy of Edge/Fog nodes has been ended by connecting the proxy servers of all blocks to a Cloud datacentre serving as the tier-0 node. To align with the characteristics of conventional network topology, the location of these computing nodes is set to be static. Furthermore, using a \textit{config.properties} file, these customised information are injected to the \textit{DataParser} class, which is easily modifiable as per the simulation use cases.
	\par Additionally, the \textit{DataParser} class provides scope for assimilating location information of multiple mobile users/IoT devices individually so that respective application services can be managed based on their distinctive mobility pattern without affecting others. Currently, two different types of mobility patterns namely \textit{DIRECTIONAL\_MOBILITY} and \textit{RANDOM\_MOBILITY} are associated with the \textit{DataParser} class through an object of \textit{Reference} class. 
	\begin{itemize}
		\item \textit{DIRECTIONAL\_MOBILITY}: This model refers to the fixed speed acyclic movement of users/IoT devices. To realise the \textit{DIRECTIONAL\_MOBILITY} model, we have at first identified a considerable number of sequential coordinates lying at the same distances across the Melbourne CBD  for a user/IoT device (as shown in Fig. \ref{Fig:mobility}.b). Later, based on those coordinates, \textit{SimEvent}s using \textit{CloudSim 5.0} are created to mimic the movement of the respective user/IoT device. During simulations, the time interval between any two of such movements is set to be equal for ensuring the fixed speed of the user/IoT device. iFogSim2 provides a scope to tune this time interval as per the test case requirements. Although this mobility model provides the pedestrial presentation of users'/IoT devices' movement, it is difficult and time-consuming to generate for each individual. Therefore, iFogSim2 also incorporates the \textit{RANDOM\_MOBILITY} model for faster generation of users'/IoT devices' movement data.     
		\item \textit{RANDOM\_MOBILITY}: There are several random mobility patterns to model the mobility behaviour of users. The \textit{RandomMobilityGenerator} class contains requirements to generate and extend different random mobility models according to various mobility characteristics, such as users' direction, speed, stopping time in each position, and users' sojourn time in the communication range of each Edge/Fog node. Currently, the \textit{RandomMobilityGenerator} class implements two well-known random mobility models, called \textit{random\_waypoint} and \textit{random\_walk} that can be used to represent the mobility model of either users or even Edge/Fog nodes, if required. Besides, in multi-user scenarios, where multiple different random mobility datasets are required, the \textit{RandomMobilityGenerator} class can be configured to generate different mobility datasets for users. Furthermore, iFogSim2 users can use the functions embedded in \textit{RandomMobilityGenerator} class to generate mobility positions in their desired Region of Interest (RoI). Fig.~\ref{Fig:mobility}.c depicts a random mobility pattern where the RoI is in Melbourne CBD.  
	\end{itemize}
	However, while parsing the location information of both Edge/Fog nodes and users/IoT devices, the \textit{DataParser} class creates separate \textit{Location} objects for each coordinate. The block-wise information of the respective entities (servers and mobile objects) can also be included in a \textit{Location} object. Furthermore, using these \textit{Location} objects, \textit{DataParser} can refer to the \textit{LocationHandler} class for sequencing the movement events of all mobile entities.    
	\begin{algorithm}[!b]
		\footnotesize
		\caption{Mobility Management Logic}\label{algo-mobility}
		\begin{algorithmic}[1]
			\Procedure{ManageMobility}{$m,t$}
			\State $dP \gets getDataParserObject()$
			\State $rF \gets getReferenceObject()$
			\State $lH \gets getLocationHandlerObject()$
			\State $L_m^t \gets dP.mobileLoction(m)$
			\State $\rho' \gets m.getParent() $
			\State $\eta \gets m.getLevel()$
			\State $\rho \gets$ null
			\State $\delta \gets rf.getMaxDistance()$
			\For{$ f_u := F_{\eta-1}$}
			\State $L_{f_u} \gets dP.fogLoction(f_u)$
			\State $\delta' \gets lH.calculateDistance(L_{f_u}, L_m^t)$
			\If{$\delta' \leq \delta$}
			\State $ \delta \gets \delta' $
			\State $ \rho \gets f_u $
			\EndIf
			\EndFor
			\If{$\rho' \neq \rho$}
			\State $ \Lambda \gets \rho'.getPlacedModulesOf(m)$
			\If{$\rho.inSameClusterOf(\rho')$}
			\State $pushModules(\rho',\rho,\Lambda)$
			\Else
			\State $ \kappa \gets$ null
			\State $ \Phi \gets getNodesInPath(\rho,$ Cloud $)$
			\State $ \Phi' \gets getNodesInPath(\rho',$ Cloud $)$
			\For{$ \varphi := \Phi$}
			\For{$ \varphi' := \Phi'$}
			\If{$\varphi = \varphi'$}
			\State $ \kappa \gets \varphi$
			\State $break$
			\EndIf
			\EndFor
			\EndFor
			\State $pushModules(\rho',\kappa,\Lambda)$
			\State $pushModules(\kappa,\rho,\Lambda)$
			\EndIf
			\State $m.setParent(\rho) $
			\State $\rho.placeModules(m,\Lambda)$
			\State $\rho'.terminateModules(m,\Lambda)$
			\EndIf
			\EndProcedure
		\end{algorithmic}
	\end{algorithm}

	\par \textit{MobilityController}: Conventionally, iFogSim requires a test script where the overall simulation environment, including the specifications of sensors, actuators, Fog nodes, and applications are defined. These classes are also set to be linked with the iFogSim simulation engine through a \textit{Controller} object. However, for simulating any mobility-driven use cases, a synthesis of such iFogSim core classes and the newly created mobility-specialised classes, including \textit{DataParser} and \textit{LocationHandler} is required. Therefore, in iFogSim2, we have created a subclass of \textit{Controller} class named \textit{MobilityController} and encapsulated the object references of those specialised classes within it so that they can collectively address the mobility-driven issues. Since the \textit{Controller} class itself is a subclass of \textit{SimEntity}, \textit{MobilityController} poses direct access to the \textit{SimEvent} class of \textit{CloudSim 5.0}. As a result, it allows the flexibility to dynamically initiate the required sequential or parallel events on different referenced objects of \textit{FogDevice} and \textit{AppModule} for mobility management.
	\par In iFogSim2, for initial placement of \textit{AppModule}s, the \textit{MobilityController} refers to an object of \textit{MobilityPlacementMobileEdgewards} class. This class replicates \textit{ModulePlacementEdgewards} of core iFogSim with an additional feature that tracks the Fog node-wise deployment of application modules/microservices for each mobile entity in the simulation environment. Once the simulation starts, \textit{MobilityController} approaches to execute events, such as launching of modules and management of resources as per the initial placement. However, when it encounters any mobility-driven event (e.g., changing of locations) as sequenced by the \textit{LocationHandler} class, \textit{MobilityController} triggers a module migration operation. In this case, \textit{MobilityController} executes the built-in \textit{\textsc{ManageMobility}} procedure as noted in Algorithm \ref{algo-mobility}. This procedure takes the object reference mobile entity $m$ and the \textit{SimEvent} triggered timestamp $t$ as the arguments (line 1) and consists of the following five phases. For a better understanding of the algorithm, Java-based object-oriented notations are used in the description. 
	\par $\bullet$ \textit{Initialisation}: In this phase, the required initialisation for the mobility management operation is performed. At first, using the \textit{Getter} methods of \textit{MobilityController} class, the object referrals of \textit{DataParser}, \textit{Reference} and \textit{LocationHandler} are extended to the procedure (lines 2-4). Later, the location $L_m^t$ of the mobile entity $m$ at timestamp $t$ is determined through the \textit{mobileLocation} method within the \textit{DataParser} class (line 5). This procedure also identifies the current upper tier contact $\rho'$ (noted as parent Fog node) of $m$ and its operating tier $\eta$ using the \textit{Getter} methods within the mobile entity (lines 6-7). An additional variable $\rho$ is also formatted in the procedure to hold the reference of $m$'s prospective new upper tier contact (due to location change) residing at the closest distance (line 8). For realising this case, another variable $\delta$ is set with a maximum distant value denoted in the \textit{Reference} class (line 9).    
	\par $\bullet$ \textit{New parent selection}: Driven by the changing of locations, this phase determines the minimum-distant new upper tier contact for the mobile entity $m$. For such an operation, Algorithm \ref{algo-mobility} primarily considers $F_{\eta-1}$ as the set of all upper tier Fog nodes corresponding to $m$ and identifies the location $L_{f_u}$ of each $f_u \in F_{\eta-1}$ with the help of \textit{DataParser} object line (lines 10-11). Later, the distance $\delta'$ from $m$ and a candidate $f_u$ is calculated. In this case, the \textit{calculateDistance} method of \textit{LocationHandler} class is exploited that takes two location objects as arguments and returns their haversine distance (line 12). Such exploration also continues for other upper tier nodes, and the Fog node that resides at a minimum distance from the mobile entity $m$ is marked as its new upper tier contact or parent node $\rho$ (lines 13-15).   
	\par $\bullet$ \textit{Intra-cluster module migration}: According to Algorithm \ref{algo-mobility}, the migration of application modules occurs when the current and new upper tier contact ($\rho'$ and $\rho$, respectively) of the mobile entity $m$ differs (line 16). To deal with such a scenario, firstly, the application modules $\Lambda$ deployed in $\rho'$ corresponding to $m$ are identified (line 17). Later, it is checked whether both $\rho'$ and $\rho$ belong to the same Fog node cluster (line 18). If this condition is satisfied, $\rho'$ simply push the modules $\Lambda$ to $\rho$ using their shared cluster communication links (line 19). Such a shifting of modules from one cluster node to another is referred to as the intra-cluster module migration. Conversely, when $\rho'$ and $\rho$ do not share a common cluster, inter-cluster migration is performed (line 20).   
	\par $\bullet$ \textit{Inter-cluster module migration}: To start this approach, Algorithm \ref{algo-mobility} firstly initialises a variable $\kappa$ which is ultimately used to refer a common accessible point for both $\rho$ and $\rho'$ (line 21). For defining $\kappa$, it also includes all Fog nodes from $\rho$ and $\rho'$ to \textit{Cloud} in separate sets $\Phi$ and $\Phi'$, respectively (lines 22-23). Later, each Fog node $\varphi \in \Phi$ and $\varphi' \in \Phi'$ are explored and mutually compared (lines 24-25). During the exploration, if any candidate $\varphi$ and $\varphi'$ indicates to the same Fog node, that node is defined to be $\kappa$, and the exploration is immediately terminated (lines 26-28). Subsequently, the application modules are pushed from $\rho'$ to $\kappa$ so that they can be further pushed to $\rho'$ and migrated across clusters (lines 29-30).                  
	\par $\bullet$ \textit{Update}: In the last phase of Algorithm \ref{algo-mobility}, necessary updates in the simulation environment based on either intra- or inter-cluster module migration are made. For example, the current upper tier contact of mobile entity $m$ is set as $\rho$ (line 31). Finally, the application modules $\Lambda$ corresponding to $m$ start execution in $\rho$ and $\rho'$ terminates them.  
	\par Algorithm \ref{algo-mobility} is a sample illustration of managing mobility in iFogSim2. From line 10 to 15 of Algorithm \ref{algo-mobility}, there are $\mathcal{O}(|F_{\eta-1}|)$ iterations, where $|F_{\eta-1}|$ denotes the number upper tier Fog nodes to the mobile entity $m$. Additionally, it has $\mathcal{O}(|\Phi| \cdot |\Phi'|)$ iterations from line 24 to 28, where $|\Phi|$ and $|\Phi'|$ define the number of nodes residing in the path from $\rho$ and $\rho'$, respectively. Such iterations helps \textit{\textsc{ManageMobility}} procedure to function with polynomial time complexity. Nevertheless, using the mobility-specialised classes and respective methods of iFogSim2, further complex and comprehensive mobility management policies can be developed. In such cases, Algorithm \ref{algo-mobility} can also be used as a benchmark.     
	\subsection{Clustering}
	In highly integrated computing environments, Edge/Fog and Cloud resources are being simultaneously considered for service delivery. Such resources are inherently heterogeneous with complementary characteristics. Distributed fog nodes usually have limited computing and storage resources compared to Cloud resources, while they can be accessed with higher bandwidth and less latency. Therefore, resource augmentation can greatly help resource-limited fog resources to be used for resource-critical applications, especially computing and storage resources \cite{goudarzi2021distributed}. Accordingly, a clustering mechanism to enable resource augmentation among Fog resources is of paramount importance. Such a clustering mechanism can also benefit multi-Cloud service providers to communicate more efficiently together. 
	\par
	The \textit{Clustering} component of iFogSim2 enables dynamic coordination and cooperation among various nodes in a distributed manner. While each node can probe and register their cluster members according to their specific clustering policy, the scheduling and other iFogSim2 features are decoupled so that clustering can be used for both centralized and distributed scenarios, such as scheduling, mobility management, and microservices.
	\par
	\textit{ClusteringController}, which is extended from \textit{Controller}, initiates the process of dynamic clustering among different nodes. In order to adapt different scenarios, \textit{ClusteringController} can trigger the clustering mechanism on various occasions, such as at the beginning of the simulation, after a specific simulation time, after a specific simulation event, or any combinations of these criteria. Nodes receiving clustering messages in \textit{FogDevice} can start their clustering process. The \textit{FogDevice} is updated with several parameters to keep the list of cluster members (CMs), bandwidth, and latency among CMs, just to mention a few. It also contains a \textit{processClustering} method which triggers the clustering process based on the policy implemented in \textit{Clustering}. As each node runs the clustering process in a distributed manner, different policies can be implemented in \textit{Clustering}.
	\par
	The current clustering policy, implemented in \textit{Clustering} class, works based on the communication range and/or latency among different nodes. In Edge/Fog computing environments, heterogeneous nodes, either wired or wireless exist. So, clustering policies can use various metrics for the creation of their clusters. The communication scope of wireless nodes is usually estimated based on their communication ranges. Therefore, for each type of Edge/Fog node, a communication range is defined according to their antenna's characteristics. Moreover, each node has a geographical position, defined in the \textit{FogDevice}. If a dataset for the position of nodes is available, their geographical position can be parsed using \textit{DataParser}. Accordingly, each node based on its geographical position and communication range can probe and create its list of CMs. Furthermore, clustering can be performed based on the average latency among each pair of nodes, regardless of whether these nodes are wireless or wired. In such scenarios, a clustering communication latency threshold is defined, through which each node can dynamically create its CMs. Algorithm~\ref{algo-clustering} represents an overview of Dynamic Distributed Clustering (DDC). First, each Edge/Fog node retrieves the information about the location of other Edge/Fog nodes (line 2). The information of nodes' positions can be obtained by each node in different ways, such as from 1) a centralized node 2) From a parent node in a hierarchical approach, or 3) GPS. Also, each Fog node is aware of the characteristics of its immediate parent, children nodes, its communication range, and acceptable communication latency (lines 3-6). Next, the latitude and longitude information of the current Edge/Fog node will be compared by all other available Edge/Fog nodes and those who are in the communication range of the current node will be added to the clustering list of the current node, $list_f^{cm}$ (lines 9-16). The \textit{calculateInRange} function is responsible to calculate the distance of the current node to other Edge/Fog nodes. The latency of the current node to each CM will be estimated and stored in \textit{mapCMToLatency} (lines 17-18). If the communication latency of CMs is a clustering factor (which is checked by the $lf$ flag), the list of current CMs will be pruned to find the CMs satisfying latency constraint (lines 19-25). Finally, the list of CMs $list^{cm}_{f}$ of the current Fog node alongside their latency mappings \textit{mapCMToLatency} will be returned as the outputs (line 26).

	\begin{algorithm}[!h]
		\footnotesize
		\caption{Dynamic Distributed Clustering (DDC) Logic}\label{algo-clustering}
		\begin{algorithmic}[1]
			\Procedure{ManageClustering}{$f,t,loc,lf$}
			\State $lHI \gets loc.locationInfo()$
			\State $\rho \gets f.getParent() $
			\State $\eta \gets \rho.getChildren()$
			\State $\delta \gets f.getRange()$
			\State $\sigma \gets f.getLatencyThresh()$
			\State $list_{f}^{cm} \gets$ null
			\State $mapCMToLatency \gets \{\}$
			\State $f_x \gets lHI.get(f).lat$
			\State $f_y \gets lHI.get(f).long$
			\For{$ f^{\prime} := \eta$} 
			\State $f^{\prime}_x \gets lHI.get(f^{\prime}).lat$
			\State $f^{\prime}_y \gets lHI.get(f^{\prime}).long$
			\State $ flag \gets calculateInRange(f_x,f_y,f^{\prime}_x,f^{\prime}_y,\delta)$
			\If{$flag$}
			\State $ list_{f}^{cm}.add(f^{\prime}) $
			\State $ latency \gets checkLatency(f,f^{\prime})$
			\State $ mapCMToLatency.get(f^{\prime}) \gets latency $
			\EndIf
			\EndFor
			\If{$lf$}
			\State $temp \gets list_{f}^{cm}$
			\For{$ f^{\prime} := list_{f}^{cm}$}
			\If{$mapCMToLatency.get(f^{\prime}) > \sigma$}
			\State $temp.remove(f^{\prime})$
			\State $mapCMToLatency.remove(f^{\prime})$
			\EndIf
			\State $list_{f}^{cm} \gets temp$
			\EndFor
			\EndIf
			
			\State return $list_{f}^{cm}, mapCMToLatency$ 
			\EndProcedure
		\end{algorithmic}
	\end{algorithm}
	
	\subsection{Microservices}
	To harvest the full potential of the Edge/Fog computing paradigm, application development has migrated from monolithic architecture towards microservice architecture. Microservices are designed as small and independent components responsible for carrying out a well-defined business function, enabling them to be moved between Edge/Fog and Cloud tiers easily \cite{joseph2019straddling,pallewatta2019microservices}. Multiple loosely coupled microservices coordinate together to build applications. Because of these characteristics, microservices can scale up and down independently based on the workload and resource availability of Edge/Fog nodes. Thus, microservice orchestration is a crucial process that combines distributed microservices to create workflows.
	
	The \textit{Microservices} component of iFogSim2 provides orchestration support to maintain seamless coordination between application microservices deployed across Edge/Fog and Cloud resources. To provide microservice orchestration, iFogSim2 models two main features: service discovery and load balancing, which help simulation of the dynamic nature of microservices within Edge/Fog computing environments.
	
	\textit{MicroserviceFogDevice}, which is created by extending \textit{FogDevice} makes it possible for Edge/Fog nodes to perform client-side service discovery and load balancing to enable decentralized orchestration among microservices. Once a request is generated in the form of a \textit{Tuple} by a consumer microservice deployed on a node, it uses \textit{ServiceDiscovery} to retrieve locations of the service provider and apply \textit{LoadBalancer} logic to determine destination node to route the created tuple. To support routing of the tuples when multiple instances of the same microservice are available on multiple Edge/Fog nodes, routing of the tuples is modelled based on the destination node id of the tuple which is set after executing the load balancer logic.
	
	\textit{LoadBalancer} and \textit{ServiceDiscovery} are initialized as members of the \textit{MicroserviceFogDevice}.
	The default implementation of the load balancer logic in iFogSim2 is based on \textit{Round Robin Load Balancing} where requests are distributed equally among microservice instances. The users of the iFogSim2 can incorporate different load balancing logic to simulate the microservice behavior by implementing \textit{LoadBalancer} interface. \textit{ServiceDiscovery} stores microservice to node mapping which can be dynamically updated at any point of time during the simulation using \textit{SimEvents}. 
	
	\textit{MicroservicesController} which is extended from \textit{Controller} initiates microservice-based application placement and orchestration. To this end, it initializes the \textit{LoadBalancer} and \textit{ServiceDiscovery} objects within each Fog node of the simulation environment and generates routing data to be used by Edge/Fog node to perform node id based routing of data tuples generated by modelled applications. Default implementation contains \textit{Shortest Path Routing} with flexibility for the user to incorporate different routing protocols. \textit{MicroservicesMobilityClusteringController} extends \textit{MicroservicesController} and integrates it with the Mobility component of iFogSim2 to provide mobility support for microservice applications. It is also integrated with the Clustering component of iFogSim2 to enable dynamic clustering among Edge/Fog nodes that host microservices. Moreover, this controller implements dynamic updating of service discovery information with user mobility-induced microservice deployment and routing data updates due to user movements.
	
	\textit{MicroservicePlacementLogic} is the base class to implement the microservice application placement policy. Users of the iFogSim2 can extend this class to implement their placement policies. As its outputs \textit{MicroservicePlacementLogic} provides two mappings:
	\begin{enumerate}
		\item \textbf{Microservice to node mapping}, which indicates where each microservice of the application gets deployed.
		\item \textbf{Service discovery information per node}, which is calculated based on the microservice to node mapping. This ensures that all nodes hosting a client microservice is aware of the locations of the service instances, that are accessed by the said microservice.
	\end{enumerate}
	\begin{algorithm}[!t]
		\footnotesize
		\caption{Scalable Microservice Placement (SMP) Logic}\label{algo-microservice}
		\begin{algorithmic}[1]
			\Procedure{ManageMicroservicePlacement}{$F,a$}
			\State $mapNodeTo\mu Inst \gets \{\}$
			\State $mapNodeToSD \gets \{\}$
			\State $m_{placed} \gets \{\}$ 
			\State $P \gets getLeafToRootPaths(F)$
			\State $mapPtoNextNode \gets getNextNode(P)$
			\State $m \gets getNextMicroservice(a,m_{placed})$
			\While{ $m$ \textbf{is not} $null$}
			\For{$p := P$}
			\State $f \gets mapPtoNextNode.get(p)$
			\If{$resources^{avail}_f \geq resources^{req}_m$}
			\State $placeModule(f,m)$
			\State $mapNodeTo\mu Inst.get(f).add(m)$
			\State $f.updateResourcesAvail()$
			\Else
			\State $notPlacedPaths.add(p)$
			\EndIf
			\EndFor
			
			\For{$p := notPlacedPaths$}
			\State $f \gets mapPtoNextNode.get(p)$
			\State $F' \gets f.getCMs()$
			\State $placed \gets \textbf{false}$
			\For{ $f' := F'$}
			\If{$resources^{avail}_{f'} \geq resources^{req}_m$}
			\State $placeModule(f',m)$
			\State $mapNodeTo\mu Inst.get(f').add(m)$
			\State $f'.updateResourcesAvail()$
			\State $placed \gets \textbf{true}$
			\State $\textbf{break}$
			\EndIf
			\EndFor
			
			\While{$placed$ \textbf{is} $false$}
			\State $f \gets getNextInPath(p,f)$ 
			\If{$resources^{avail}_f \geq resources^{req}_m$}
			\State $placeModule(f,m)$
			\State $mapNodeTo\mu Inst.get(f).add(m)$
			\State $f.updateResourcesAvail()$
			\State $placed \gets \textbf{true}$
			\State $mapPtoNextNode.get(p).set(f)$
			\EndIf
			\EndWhile
			\EndFor
			\State $m_{placed}.add(m)$
			\State $m \gets getNextMicroservice(a,m_{placed})$
			\EndWhile
			\State $mapNodeToSD \gets generateSD(mapNodeTo\mu Inst)$
			\State return $mapNodeTo\mu Inst, mapNodeToSD$ 
			\EndProcedure
		\end{algorithmic}
	\end{algorithm}
	
	Algorithm \ref{algo-microservice} provides an overview of Scalable Microservice Placement Logic (SMP), which is the default microservice placement policy available in iFogSim2. It is an edgeward placement algorithm for microservices, which focuses on horizontally scaling microservices among Edge/Fog nodes of the same cluster before moving towards upper-tier nodes of the Edge/Fog hierarchy. First, the placement policy identifies leaf to root paths ($P$) considered for placement (line 5) and initializes the \textit{mapPtoNextNode} with the first eligible node in each path for the placement process (line 6). Leaf to root paths are calculated based on the physical topology created by all available Edge/Fog nodes ($F$). Each path in $P$ starts with a user/IoT device and traverses upward within the Edge/Fog hierarchy until it reaches the Cloud. For the microservice application, the next eligible microservice for placement is determined by traversing its DAG representation (line 7). A microservice becomes eligible for placement if all predecessor microservices are mapped to nodes. Afterward, the policy iteratively tries to place eligible microservices onto the next eligible node of each path (lines 9-16). If sufficient resources are not available within the considered node, the policy considers cluster members (lines 21-27) before moving onto the next tier (lines 28-35). After all microservice instances are mapped to nodes, the algorithm generates service discovery information for each node hosting client microservices (line 38). 
	\par
	These objects together create a platform to model microservices in Edge/Fog computing environments, while capturing their dynamic, independent, and scalable nature.
	
	\section{Performance Evaluation}
	\label{sec_performance}
	This section discusses the simulation of a set of Edge/ environments using iFogSim2 for different application case studies, including Audio Translation Service (ATS), Cardiovascular Health Monitoring (CHM), and Crowd-sensed Data Collection (CDC). Then, we evaluated the efficiency of various combinations of iFogSim2's built-in Mobility, Clustering, and Microservice management policies with respect to latency, network usage, and energy consumption for each case study. We also parameterised the lightweight and modular architecture of iFogSim2 in terms of RAM usage and execution time and compared it with the existing simulators, including IoTSim-Edge \cite{Jha} and PureEdgeSim \cite{Mechalikh}. The use cases and the experiment results are discussed below.  
	
	\subsection{Case study 1: Audio Translation Service (ATS)}
		\begin{figure*}[!ht]
		\centering 
		\includegraphics[width=0.90\textwidth]{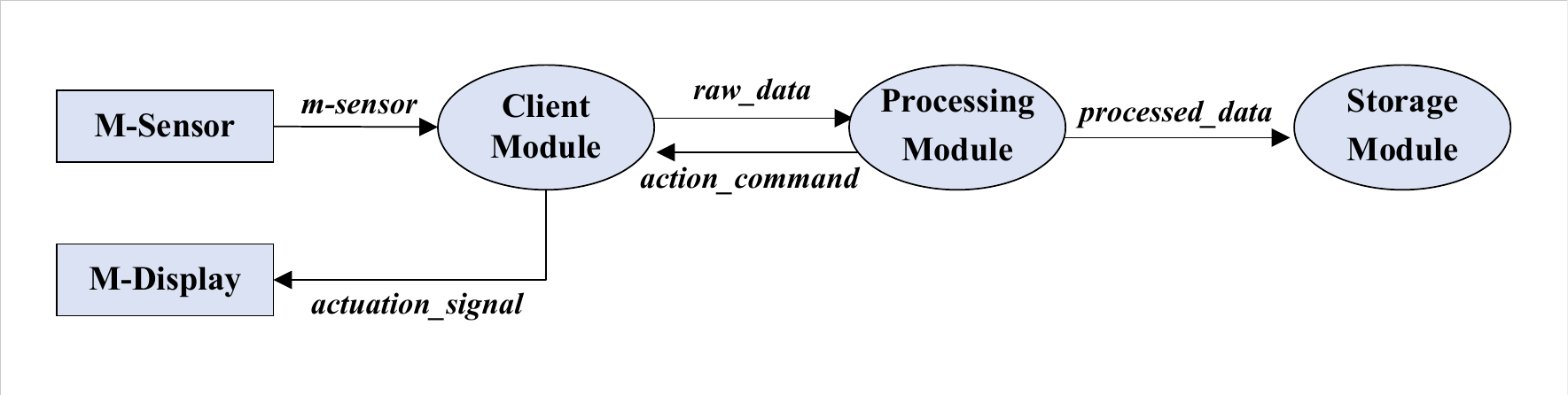}
		\caption{Application model for the Audio Translation Service (ATS)}
		\label{Fig:transApp}
	\end{figure*}
	Translation service is highly recommended for tourists, especially when they are visiting non-native language speaking countries. Currently, Google and Microsoft offer different translator services to the users, which mainly deal with text and imagery inputs \cite{transService}. Since the frequency and variations of such inputs can be easily estimated or controlled, their processing is usually performed by following a specific set of operations without requiring any additional services. As a result, most of state-of-the-art smartphones can execute these translation services with the available computing resources they have. However, for audio-based translation, various computation-intensive data pre-processing operations are required as the pitch intensity varies between the users, and the background noises always couple tightly with the actual data \cite{ATS}. Conversely, for smartphones, the real-time adjustment or update of external services for performing these operations is not often feasible due to additional overhead. In such scenarios, the exploitation of Fog computation can be a potential solution for Audio Translation Service (ATS). 
	\par However, in any Fog computing-based ATS system, a majority of users is expected to be mobile and their smartphones are regarded as the data sources. Therefore, to meet the desired QoS, efficient mobility-aware service management techniques are required for such an ATS. Considering this issue, we have modelled a mobility-driven simulation case study on Fog computing-based ATS in iFogSim2. The details of the application model, simulation parameters, comparing mobility management policies and their performances for this case study are discussed below.             
	%

	%
	
	\subsubsection{Application Model}
	To align with the distributed data flow approach adopted by core iFogSim, we have modelled the application for ATS as a Directed Acyclic Graph (DAG) (shown in Fig. \ref{Fig:transApp}). It consists of three application modules, which are described in the following
	\begin{itemize}
		\item Client module: It is deployed on smartphones that primarily grasp audio data from the integrated sensors. The Client module also performs necessary authentications to access the ATS and forwards the data to the Processing module for further analysis.
		
		\item Processing module: It is expected to execute by the tier-2 nodes for faster interactions with the smartphones. However, various computation-intensive Artificial Intelligence (AI)-enabled audio data analysis operations including data filtration, noise reduction, pitch classification, and speech segmentation are performed by the Processing module. The results of these analyses are then pushed back to the Client module so that they can be displayed to the user via the smartphone display.
		
		\item Storage module: The Processing module forwards the input data and analytical outputs to the Storage module for periodic updates of the AI models and thus ensures the enhanced performance of the ATS.
	\end{itemize}
	
	Considering the amount of audio data generated through such an ATS, it is recommended to host the Storage module in Cloud for further scalability.

	\subsubsection{Simulation Environment}
	The simulation environment for the ATS use case is made highly aligned with the EUA dataset of iFogSim2 having 118 Fog gateways residing at 12 different blocks across the Melbourne CBD. We assume that the smartphones of mobile users can connect with any of the gateways (tier-2 nodes). The gateways of a particular block can also interact with a Cloud datacentre (tier-0 nodes) via a proxy server (tier-1 nodes). The specifications of the computing infrastructure along with that of application modules are presented in Table \ref{Tab_system}. The simulation experiments are conducted on an \textit{Intel Core 2 Duo CPU @ 2.33-GHz with 2GB-RAM configured computer}, and the fractional selectivity of input-output relationship within a module is set to be 1.0. The numeric values of the simulation parameters have been extracted from the existing literature as mentioned in \cite{performanceParam1,performanceParam2}.  
	\begin{table*}[!t] 
		\caption{Simulation parameter for the ATS}\label{Tab_system} 
		\centering
		\scriptsize
		\begin{tabular}{|p{2 cm}p{2 cm}p{2 cm}p{2 cm}p{2 cm}|}
			\hline
			\multicolumn{5} {|l|}{Duration of experiment : 500 seconds}\\\hline
			\multicolumn{5} {|l|}{Number of location change events:140}\\\hline
			Resource type $\Rightarrow$ \newline Configuration $\Downarrow$  & Cloud VM & Proxy server & Fog gateway & Smart-phone \\ 
			Numbers & 10 & 12 &  118 &  1 \\
			Speed (MIPS) & 4480 & 3600-4000 &  2800-3000 &  500 \\ 
			RAM (GB) & 16 & 16 & 8  & 1  \\
			Uplink (MBPS) & 100 & 10 & 50 & 100 \\
			Downlink (MBPS) & 100 & 20 & 100 & 200  \\
			Busy power (MJ) & 1468 & 428 & 206 & 60  \\
			Idle power (MJ) & 1332 & 333 & 170 & 35 \\\hline
			Attribute $\Rightarrow$ \newline Module $\Downarrow$  & RAM (GB) & Input (MB) & Output (MB) & CPU length (MI) \\
			Client & 0.10 & 2 &  2.5 &  500 \\
			Processing & 4 & 2.5  & 1.5  & 2500 \\
			Storage & 4 & 1 & 1 & 1000 \\\hline
		\end{tabular} 
	\end{table*}
	\subsubsection{Comparing Policies}
	While imitating the ATS case study in iFogSim2, the movement of smart-phones are set to vary using both directional and random mobility pattern. Furthermore, we have used three different mobility management techniques in the simulated Fog computing environment to deal with such movement. These techniques are listed below.    
	\begin{itemize}
		\item  \textbf{Cloud-centric migration}: In this approach, the current upper tier contact (source gateway) of a mobile smartphone pushes the respective application modules directly to the Cloud VMs. Later, the Cloud VMs pushes the modules to the new upper tier contact (destination gateway) of the smartphone.   
		\item \textbf{Non-hierarchical migration}: By connecting all Fog gateways through a mesh communication channel, this approach allows the direct migration of application modules between source and destination. As a result, the upper tier Fog nodes remain uninvolved during module migration, leading it to be a non-hierarchical operation. 
		\item \textbf{Intra/Inter-cluster migration}: This approach refers to the built-in mobility management policy of iFogSim2 as discussed in Algorithm \ref{algo-mobility}. It only involves the upper tier Fog nodes in migrating modules if the source and destination gateway do not belong to the same cluster. Here, the node clustering is performed by Algorithm \ref{algo-clustering}.  
	\end{itemize}
	\subsubsection{Results}
	The performance of the comparing techniques is discussed below. 
	\par
\begin{figure}[!t]
		\centering 
		\includegraphics[width=\columnwidth]{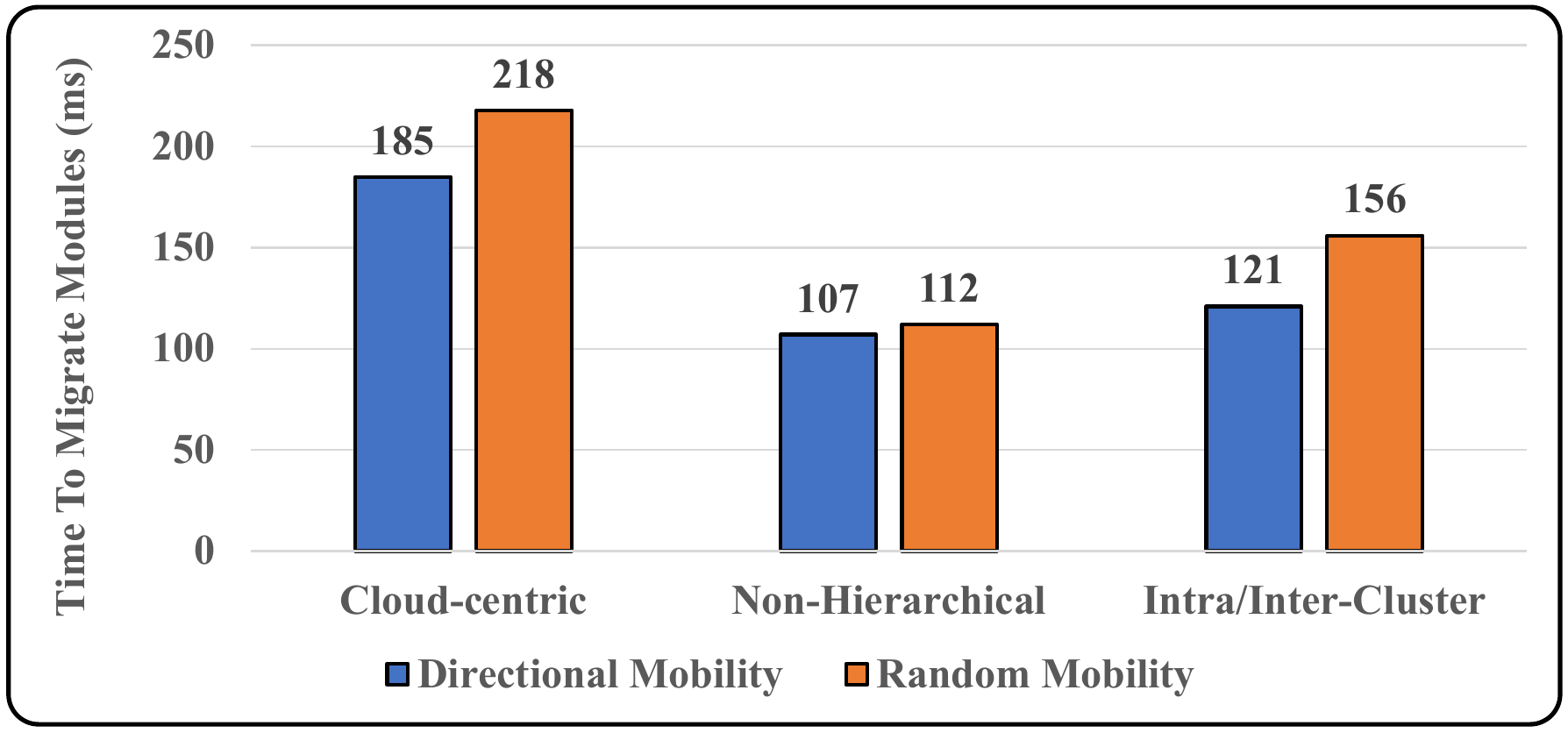}
		\caption{Time to migrate application modules}
		\label{Fig:migTime}
	\end{figure}

$\bullet$\textbf{ Migration time}: Fig. \ref{Fig:migTime} depicts the delay in migrating application modules for different comparing mobility management policies. Since Cloud datacentres reside at a multihop distance from the gateways, the transfer of application modules to the Cloud and later their forwarding to the destination gateway increases the overall migration delay for the Cloud-centric approach. Conversely, the Intra/Inter-cluster migration technique exploits all possible options to select a common accessible node (CAN) for both source and destination gateway not only in the node hierarchy but also in horizontal levels. As a result, it is more likely to find a CAN in the proximity of the gateways than that of a Cloud-centric approach, minimising the migration delay. However, the Non-hierarchical policy provides the most desirable outcome in this case as it exploits the mesh connectivity between source and destination gateway during module migration reducing the delay significantly. Nevertheless, as the assurance of large-scale mesh connectivity across the gateway is costly, such an approach is feasible only when user mobility is confined. 
	\par Furthermore, directional mobility presents better management of migration delay than random mobility in all comparing policies. It happens because, in directional mobility, user speed remains the same. As a result, despite location change, their source and destination gateway are less likely to vary within a short distance. Consequently, it reduces the number of total migration events and lessens the delay. Conversely, during random mobility, the number of migration events can increase unevenly, resulting in increased migration delay.
	\par
\begin{figure}[!t]
		\centering 
		\includegraphics[width=\columnwidth]{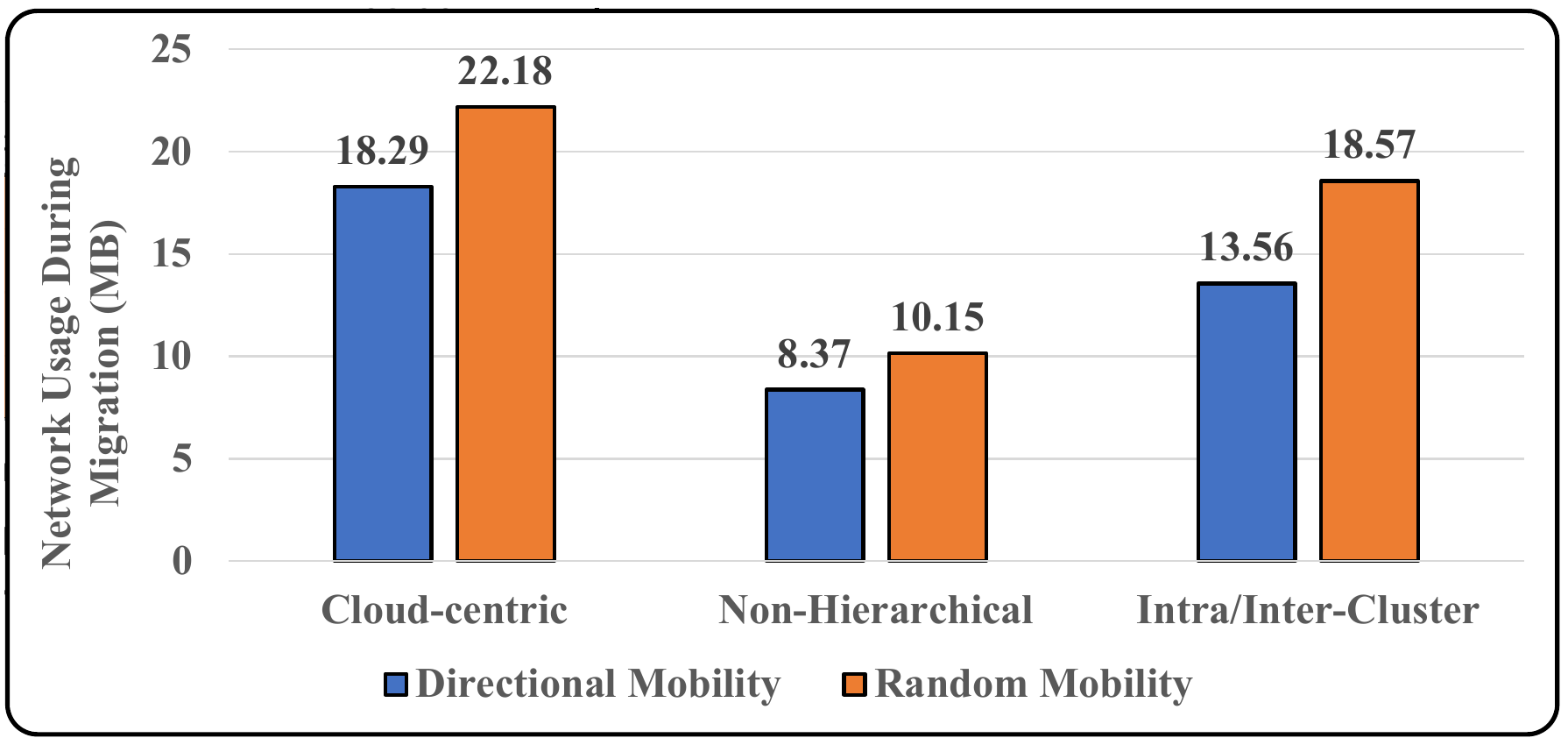}
		\caption{Network usage during module migration}
		\label{Fig:netUsage}
	\end{figure}

$\bullet$\textbf{ Network usage}: Fig. \ref{Fig:netUsage} portrays the network resource usage during module migration for different simulating techniques. As the realisation of the Cloud-centric approach involves multiple nodes within the communication path between gateways and Cloud to migrate modules in both directions, it consequently increases their collective network usage. Nevertheless, the Intra/Inter-cluster technique performs slightly better in this case as it attempts to reduce the involvement of intermediate nodes during module migration and consequently lessens their network usage. Conversely, the Non-hierarchical approach only exploits the network resources which is available within the communication link between the source and destination gateway, resulting in the least usage. 
	\par Moreover, as the number of migrating events increases with random mobility, network usage increases. On the other hand, by limiting the occurrence of such events, directional mobility can help to lower network resource usage.
	\par
\begin{figure}[!ht]
	\centering 
	\includegraphics[width=\columnwidth]{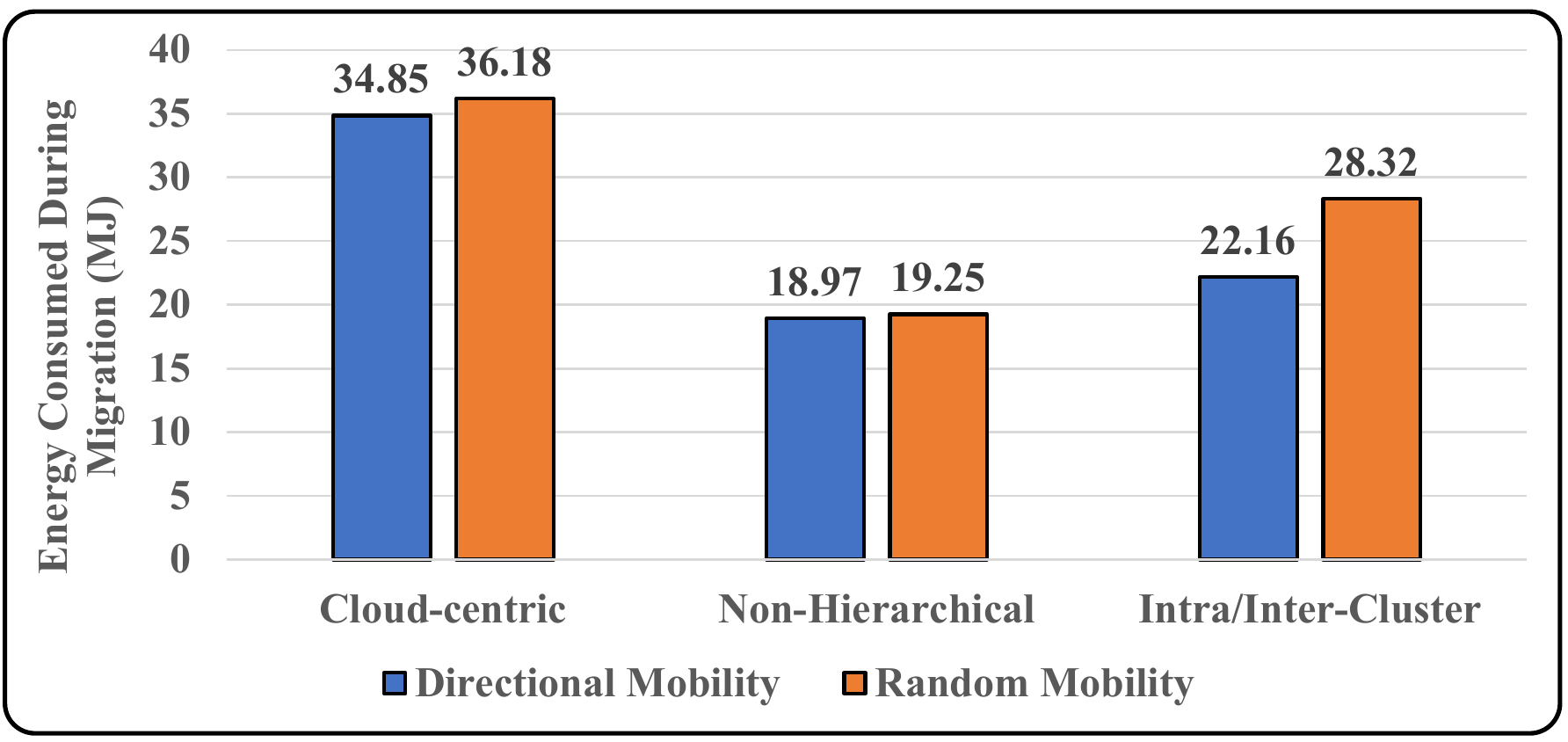}
	\caption{Energy consumed during module migration}
	\label{Fig:enerConsume}
\end{figure}

$\bullet$\textbf{ Energy consumption}: Fig. \ref{Fig:enerConsume} illustrates the consumption of energy during module migration for different adopted policies. For the Cloud-centric approach, the increment in energy usage is obvious as it directly involves Cloud datacentres conventionally consuming a significant portion of energy in the distributed computing ecosystem. The involvement of other intermediate nodes further contributes to increasing the overall energy consumption. The random mobility of users can also elevate energy usage during mobility management by increasing the migration frequency. In comparison, both Intra/Inter-cluster and Non-hierarchical module migration approaches perform well in managing energy usage as they resist the involvement of Cloud and intermediate nodes to a greater extent for such an operation.
	
	\begin{figure*}[!h]
	\centering 
	\includegraphics[width=0.9\textwidth]{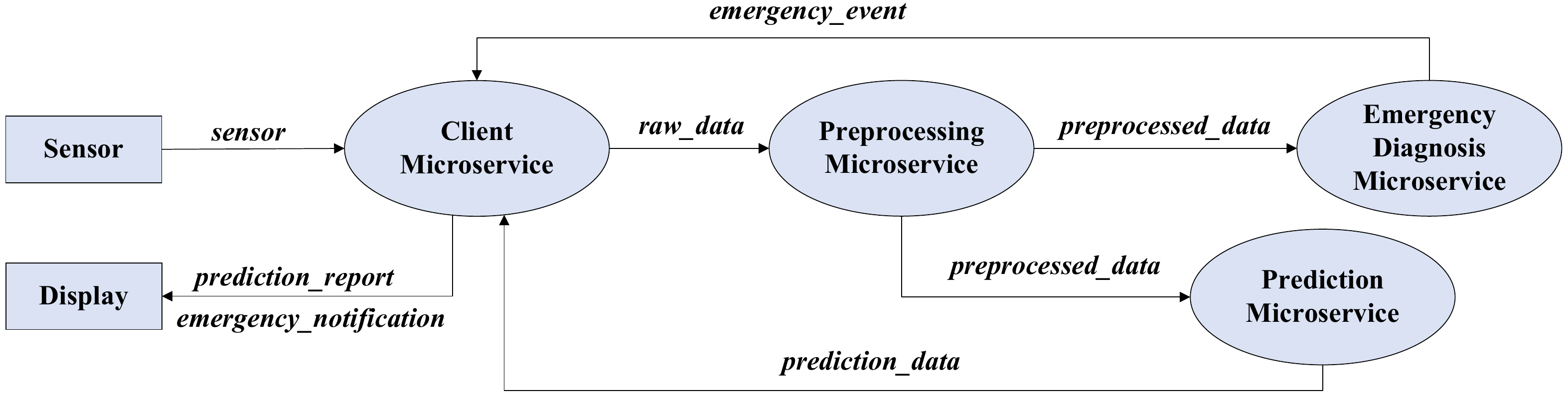}
	\caption{Microservice application model for the Cardiovascular Health Monitoring (CHM)}
	\label{Fig:ecgApp}
\end{figure*}
	\subsection{Case study 2: Cardiovascular Health Monitoring (CHM)}
	Electrocardiogram (ECG) monitoring is a widely used method for diagnosing heart diseases using both reactive and proactive methods \cite{ECG-Omar,pallewatta2019microservices}. In IoT-based smart healthcare, wearable sensors are used to sense and transmit ECG signals towards analysis platforms that host applications developed for detecting concerning heart conditions. Such applications are designed to perform multiple tasks, including: filtering ECG data to remove data anomalies, ECG feature extraction and generating emergency warnings in real-time, long-term collection and analysis of data to make predictions to ensure preventive measures \cite{goudarzi2021distributed}. 
	\par
	Design and development of such IoT applications increasingly use modular architectures, especially microservices architecture to enable balanced deployment of real-time tasks within resource-constrained Fog resources and latency tolerant tasks within Cloud datacentres. Hence, Cardiovascular Health Monitoring (CHM) is designed following microservice architecture and modelled using DAG-based modeling of applications integrated into iFogSim2.
	\par
	\subsubsection{Application model}
	Fig \ref{Fig:ecgApp} shows the microservice architecture of the CHM application, where vertices represent each microservice and edges depict the data dependencies among microservices. CHM consists of four microservices, namely \textit{Preprocessing Microservice}, \textit{Emergency Diagnosis Microservice}, \textit{Prediction Microservice}, and \textit{Client Microservice}. The specifications of each microservice are described in what follows: 
	\begin{itemize}
		\item Client Microservice: This is the mobile front end of the CHM application. Client microservice is deployed on users' smartphones and receives raw ECG signals transmitted by the sensors that are wirelessly connected to each smartphone. Also, it is responsible for sending sensor data towards Preprocessing Microservice placed in either Edge/Fog or Cloud and displaying results received after processing.
		\item Preprocessing Microservice:  The Preprocessing microservice performs data cleaning using filters to filter out noise added to ECG sensor data during transmission. Moreover, data anomalies in the sensed data are also removed before sending data for further processing.
		\item Emergency Diagnosis Microservice: This microservice is responsible for real-time analysis and identification of concerning health conditions like heart attacks and sending back a warning signal towards the client microservice to trigger an emergency notification.
		\item Prediction Microservice: Prediction microservice stores and analyses ECG time series data using machine learning models to predict health risks to the patients. Prediction reports are sent back to the mobile front end to be displayed for users. 
	\end{itemize}
	These microservices communicate together to monitor and predict the cardiovascular health of users. Preprocessing and  Emergency Diagnosis microservices form a latency-critical service to be placed on Fog or Cloud, based on the placement policy whereas Prediction Microservice represents a service that requires high computation and storage resources and is expected to be placed in the Cloud.

	\begin{table*}[!t] 
		\caption{Simulation parameter for the CHM}\label{Tab:CHM_topology} 
		\centering
		\scriptsize
		\begin{tabular}{|p{2 cm}p{2 cm}p{2 cm}p{2 cm}p{2 cm}|}
			\hline
			\multicolumn{5} {|l|}{Duration of experiment : 20000 seconds}\\\hline
			\multicolumn{5} {|l|}{Number of location change events:140}\\\hline
			Resource type $\Rightarrow$ \newline Configuration $\Downarrow$  & Cloud VM & Proxy server & Fog gateway & Smart-phone \\ 
			Numbers & 16 & 1 &  6 &  25 \\
			Speed (MIPS) & 2500-3000 & 2500-3000 &  2500-3000 &  500 \\ 
			RAM (GB) & 16 & 16 & 8  & 1  \\
			Uplink (MBPS) & 100 & 10 & 50 & 100 \\
			Downlink (MBPS) & 100 & 20 & 100 & 200  \\
			Busy power (MJ) & 107.339 & 107.339 & 107.339 & 87.530  \\
			Idle power (MJ) & 83.433 & 83.433 & 83.433 & 82.440 \\\hline
			Attribute $\Rightarrow$ \newline Module $\Downarrow$  & RAM (GB) & Input (MB) & Output (MB) & CPU length (MI) \\ 
			Client & 0.10 & 0.5 &  0.5 &  1000 \\
			Preprocessing & 0.5 & 0.5  & 0.5  & 2000 \\
			Emergency Diagnosis & 0.5 & 0.5 & 0.5 & 2500 \\
			Prediction & 2 & 0.5 & 0.5 & 4000 \\
			\hline
		\end{tabular} 
	\end{table*}
	
	\subsubsection{Simulation Environment}
	For this case study, a physical topology of 7 Fog nodes is used, which consists of 6 Wifi gateways (tier-2 nodes) connecting to a single proxy server. Besides, the proxy server (tier-1 nodes) is connected to the Cloud datacentre (tier-0 nodes). Also, 25 smartphones with randomly generated locations connect with the Wifi gateways to send ECG sensor data towards the Fog environment. The specifications of the computing infrastructure along with that of application modules are presented in Table~\ref{Tab:CHM_topology}. Furthermore, to model the physical topology, \textit{MicroserviceFogDevcie}, \textit{Sensor} and \textit{Actuator} classes of iFogSim2 are used. The simulation experiments are conducted on an \textit{Intel Core i7 CPU @ 1.80-GHz with a 4GB-RAM computer} and the fractional selectivity of the input-output relationship within a module is set to be 1.0. The numeric values of the simulation parameters have been extracted from the existing literature as mentioned in \cite{pallewatta2019microservices,goudarzi2020application,xu2019computation}.
		\begin{figure}[!ht]
		\centering 
		\includegraphics[width=\columnwidth]{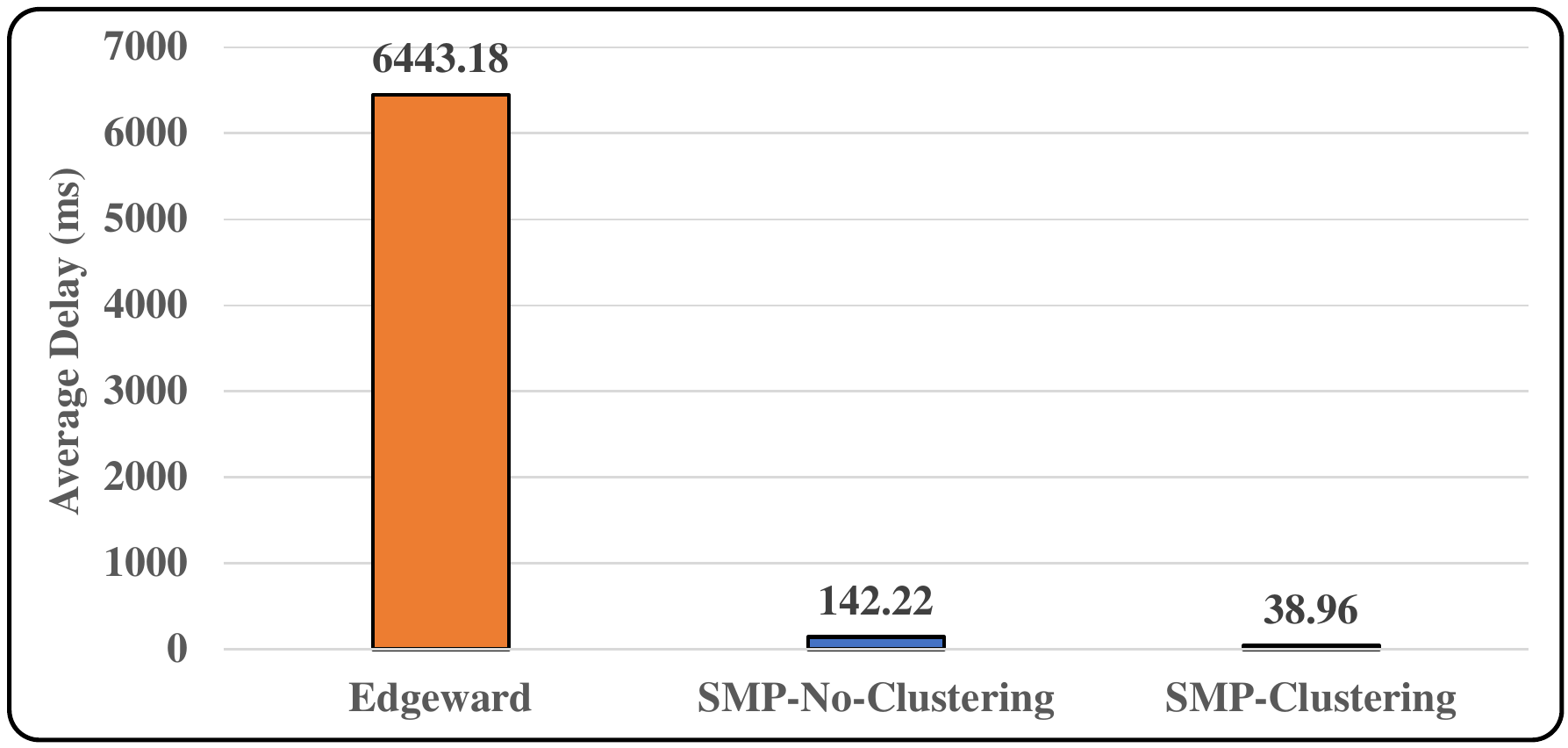}
		\caption{The average delay of the CHM application}
		\label{Fig:CHM_latency}
	\end{figure}
	\subsubsection{Comparing Policies}
	%
	For the experiments, three different placement techniques are used, as follows:
	\begin{itemize}
		\item \textbf{Edgeward:} This placement technique only considers vertical scalability of application modules without taking Fog node clustering and horizontal scalability-based load balancing into consideration.
		\item \textbf{SMP-No-Clustering:} It uses horizontal scalability and load balancing features available in microservice orchestration of iFogSim2 but does not implement clustering of Fog nodes.
		\item \textbf{SMP-Clustering:} Such an approach makes use of both microservice orchestration and clustering features available in iFogSim2.
	\end{itemize}
	
	\subsubsection{Results} The performance of the comparing techniques are discussed below.
	
	\par$\bullet$\textbf{ Average delay of control loop:}
	The control loop of the CHM application, which is the most latency-sensitive loop, consists of \textit{Client microservice  $\rightarrow$ Preprocessing Microservice $\rightarrow$  Emergency Diagnosis Microservice $\rightarrow$  Client microservice}. The less average delay for this control loop demonstrates better placement decision and coordination among computational resources. Fig.~\ref{Fig:CHM_latency} depicts the average delay for the execution of this control loop for three placement techniques. It depicts that the average execution delay of the control loop significantly decreases for SMP-No-Clustering and SMP-Clustering in comparison to the Edgeward. Edgeward placement moves microservices upwards to the Fog hierarchy, such that a single instance of each microservice is deployed for each edge to Cloud path. Due to the resource-constrained nature of Fog nodes, this approach places microservice instances in higher Fog tiers, thus increasing average delay. Also, the results show that the SMP-Clustering outperforms SMP-No-Clustering. Even though SMP-No-Clustering uses horizontal scalability, due to lack of clustering, microservices are scaled among nodes on multiple Fog tiers whereas, in the case of SMP-Clustering, clusters are dynamically formed among Fog nodes of the same hierarchical tier, which enables microservices to be horizontally scaled within nodes of the same tier before moving to the upper tier. Hence, the SMP-Clustering scenario places latency-critical microservices closer to the edge network which results in lower average delay.
	
	\par$\bullet$\textbf{ Energy consumption:}
	\begin{figure*}[!htbp]
		\centering
		\begin{minipage}[t]{0.49\textwidth}
			\centering
			\vspace{0pt}
			\includegraphics[width=\textwidth,align=t]{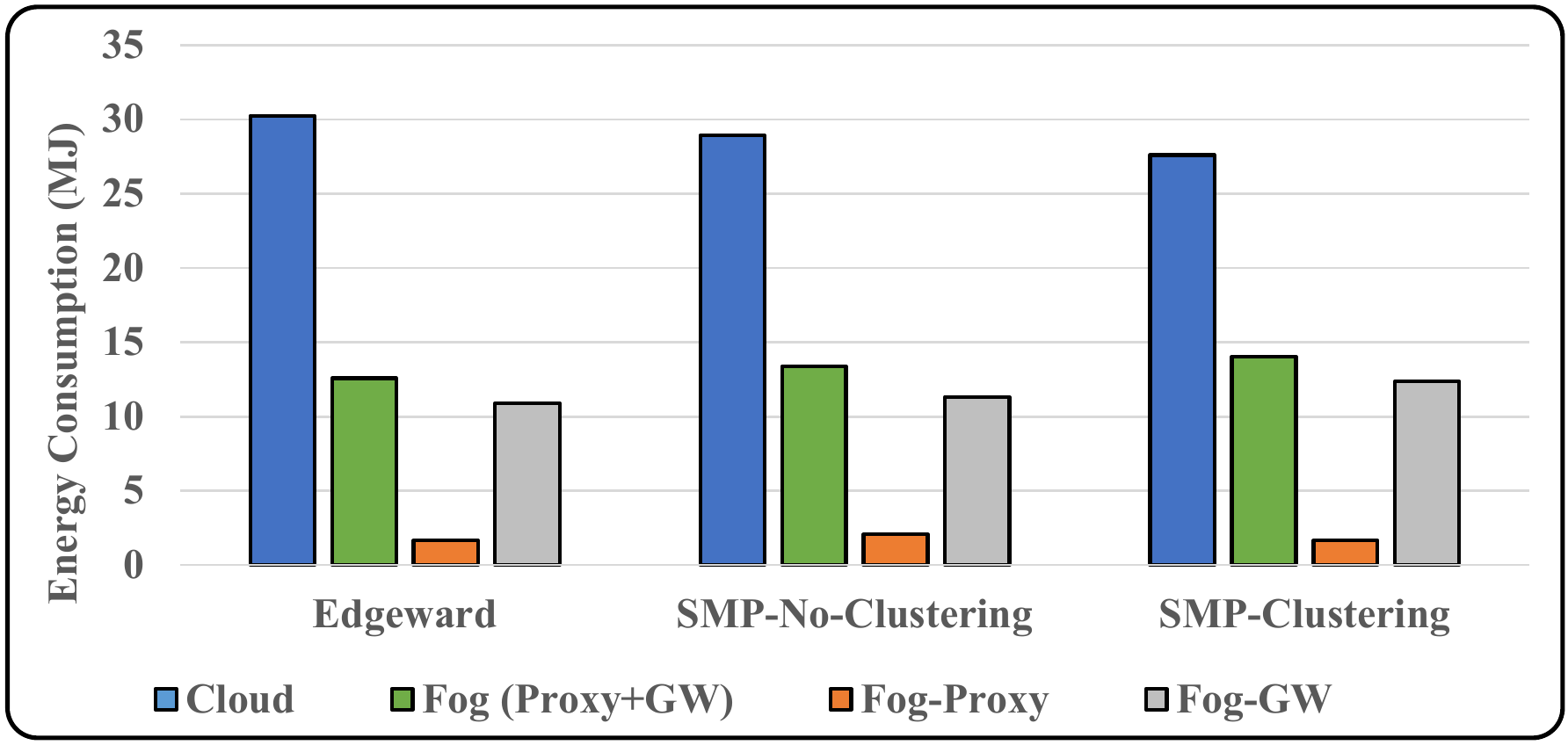}
		\end{minipage}
		\begin{minipage}[t]{0.49\textwidth}
			\centering
			\vspace{0pt}
			\includegraphics[width=\textwidth,align=t]{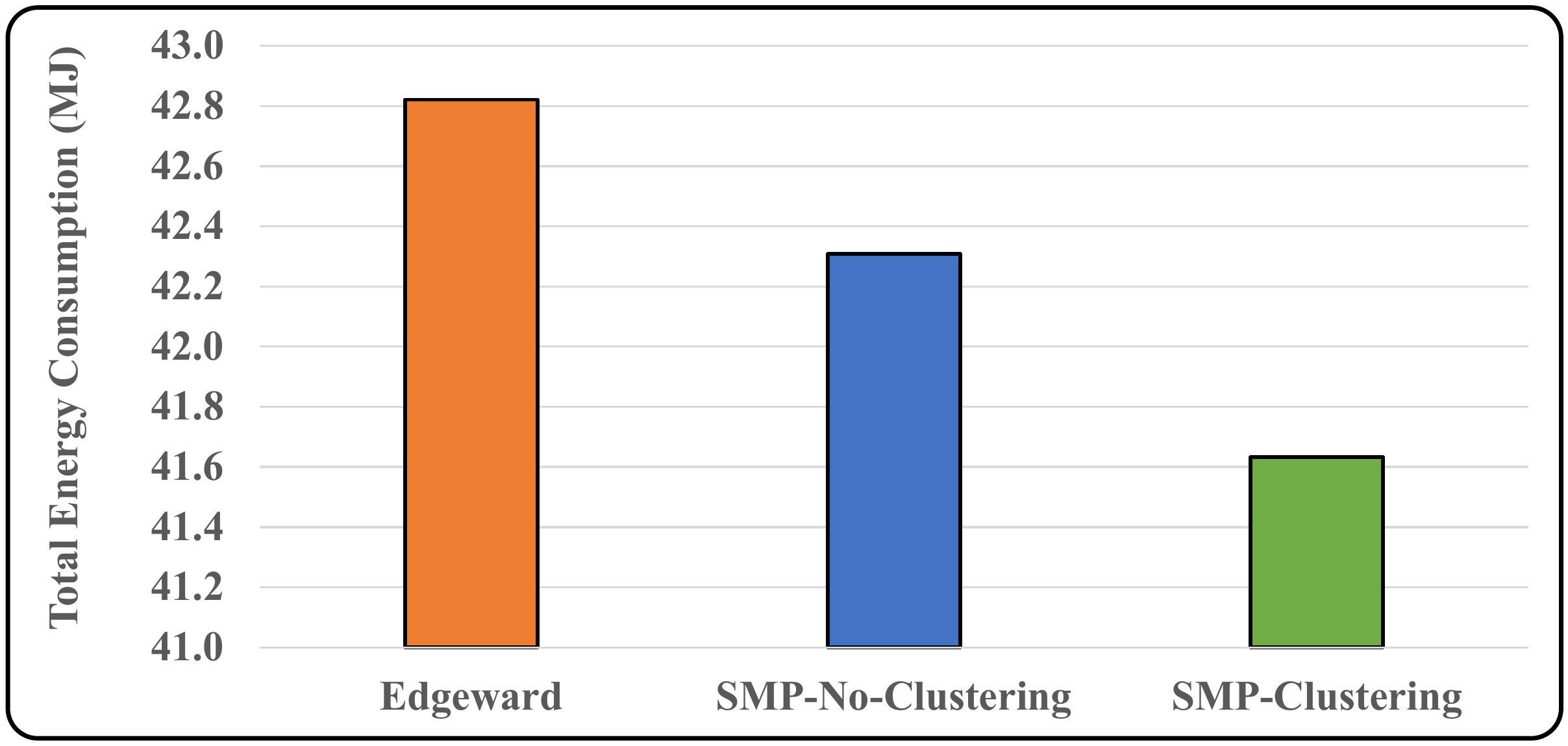}
		\end{minipage}
		\footnotesize{(a)\hspace{250pt}(b)}\\
		\caption{Energy consumption for running the CHM application (a) Energy consumed by the resources of different tiers, (b) Total energy consumption of resources}
		\label{Fig:CHM_Energy}
	\end{figure*}
	Figure~\ref{Fig:CHM_Energy} shows the amount of energy consumed by different classes of nodes and total energy consumption obtained through each technique. As Fig.~\ref{Fig:CHM_Energy}.a demonstrates, the energy consumption of Cloud resources is higher in the Edgeward technique because it places most of the microservices on the Cloud VMs. However, the energy consumption of Cloud resources for SMP-No-Clustering and SMP-Clustering are lower while they consume more energy in the Fog tier due to running more microservices in the resources of that tier. While energy consumption at different tiers depends on the number of microservices running on that tier, Fig.~\ref{Fig:CHM_Energy}.b depicts that the total energy consumption of all nodes is reduced by in SMP-No-Clustering and SMP-Clustering. It highlights the positive effect of more efficient distribution, scaling, and load balancing of microservices in the Fog tier compared to more centralized approaches. Also, the results of SMP-Clustering prove the potential of clustering for better scaling and load balancing of microservices either vertically or horizontally.
	
	\par$\bullet$\textbf{ Network usage:}
	\begin{figure}[!t]
		\centering 
		\includegraphics[width=\columnwidth]{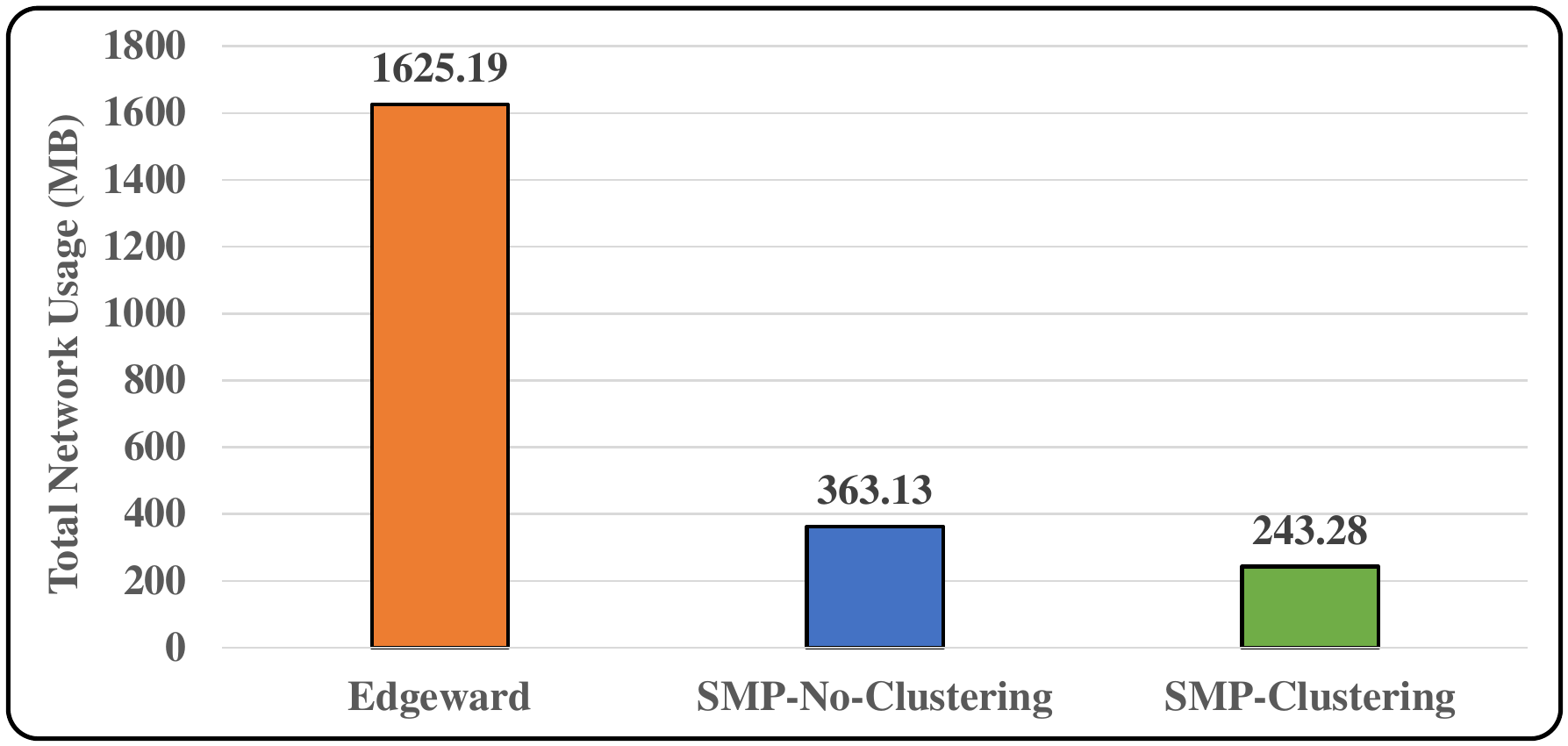}
		\caption{The total network usage of the CHM application}
		\label{Fig:CHM_nw}
	\end{figure}
	The total amount of data transferred in the network is an important metric for the evaluation of different techniques. Techniques resulting in high data transmission may lead to congestion in the network, service interruption, or increasing average delay of the applications' control loop, especially in high-density networks. Fig.~\ref{Fig:CHM_nw} illustrates the total network usage of different techniques in Megabytes (MB). It demonstrates that Edgeward placement incurs higher network usage due to excessive usage of higher tier Fog nodes and Cloud datacentre in comparison to the other two techniques. Furthermore, it shows that SMP-Clustering results in lower network usage compared to SMP-No-Clustering. Although the clustering mechanism embedded in the SMP-Clustering requires data transmission for the formation of clusters, it is a lightweight mechanism in terms of network usage. Hence, in a long simulation time, SMP-Clustering outperforms other techniques due to the efficient usage of the lower tier Fog nodes and better load balancing.

	\subsection{Case study 3: Crowd-sensed Data Collection (CDC)}
	Crowd-sensing exploits internet-connected sensors to collect vast amounts of data that can be analysed to retrieve complex information. Crowd-sensed Data Collection (CDC) application represents a mobile crowd-sensed scenario that aids urban road network planning. Within urban settings, road system design and traffic signal controlling are extremely challenging. Thus, these tasks can benefit from complex machine learning algorithms. As such algorithms require large amounts of data for accurate decision making, vehicular crowd-sensing is used as a solution for data collection. Sensors onboard mobile vehicles sense and transmit real-time location and speed data that can be used to derive traffic conditions of the road networks.  Using this method, any vehicle can voluntarily share data with the data analytic platforms, which results in a collection of large volumes of data. Such applications can benefit from Fog computing environments to process the data closer to the edge network, thereby reducing the burden on the data transmission networks connecting sensors to the Cloud. So we design a CDC application following the microservice architecture and modelled it using DAG-based  modeling of applications integrated with iFogSim2.
	
	\subsubsection{Application model}
\begin{figure*}[!t]
		\centering 
		\includegraphics[width=0.9\textwidth]{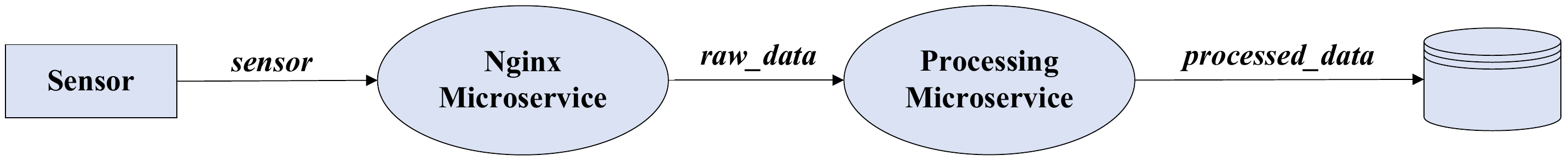}
		\caption{Microservice application model for the Crowd-sensed Data Collection (CDC)}
		\label{Fig:CDCapp}
	\end{figure*}

	Fig \ref{Fig:CDCapp} shows the microservice architecture of the CDC application, where vertices represent each microservice and edges depict the data dependencies among microservices. CDC consists of two microservices, namely \textit{Nginx Microservice}, \textit{Processing Microservice} and a database to store data for further processing. The specifications of each microservice are described in what follows: 
	\begin{itemize}
		\item Nginx Microservice: This is the webserver that acts as the gateway to the processing microservice. Nginx microservice receives data, which is generated by the vehicular sensor network, and routes that data towards the Processing microservice to perform data analytics. Also, it is responsible for load balancing requests among multiple Processing microservices.
		\item Processing Microservice: The processing microservice is responsible for sanitizing the sensor data, extracting features that represent trajectories of vehicles, and sending the processed data towards the Cloud to be saved in a time-series database. Data analytic platforms can use crowd-sensed data stored in the database for urban planning.
	\end{itemize}
	
	\begin{table*}[!h] 
		\caption{Simulation parameter for the CDC}\label{Tab_system3} 
		\centering
		\scriptsize
		\begin{tabular}{|p{3 cm}p{2 cm}p{2 cm}p{2 cm}p{2 cm}|}
			\hline
			\multicolumn{5} {|l|}{Duration of experiment : 500 seconds}\\\hline
			\multicolumn{5} {|l|}{Tuple generation rate : 5/seconds}\\\hline
			\multicolumn{5} {|l|}{Mobility interval : 10-50 seconds}\\\hline
			Resource type $\Rightarrow$ \newline Configuration $\Downarrow$  & Mobile vehicles & Tier-2 Node & Tier-1 Node & Tier-0 Node \\ 
			Percentage of nodes & 30\% & 30\% & 20\% & 20\% \\
			Speed (MIPS) & 500-1000 & 2000-2500 &  3000-2500 & 4000-5000 \\ 
			RAM (GB) & 2 & 4 & 8  & 16  \\
			Uplink (MBPS) & 100 & 50 & 10 & 100 \\
			Downlink (MBPS) & 200 & 100 & 50 & 150  \\
			Busy power (MJ) & 50-100 & 200-300 & 400-600 & 1500-2000  \\
			Idle power (MJ) & 20-30 & 80-100 & 150-200 & 700-900 \\\hline
		\end{tabular} 
	\end{table*}
	
	\subsubsection{Simulation environment}
	Table \ref{Tab_system3} presents the specifications of general simulation parameters used in imitating the CDC case study. The value of simulation parameters within a specific range is
	determined by a pseudo-random number generator. Moreover, the computing environment is set to be hierarchical having mobile vehicles at the lowest tier. Tier-2 Fog nodes are marked as the gateway followed by proxy servers and Cloud at tier-1 and tier-0, respectively. An \textit{Intel Core 2 Duo CPU @ 2.33-GHz with 2GB-RAM configured computer} has been used to execute the simulation script and perform the experiments. The numeric values of the simulation parameters have been extracted from the existing literature as mentioned in \cite{pallewatta2019microservices,performanceParam1}.      
	\subsubsection{Comparing Simulators}
	The simulation environment for the CDC use case has been implemented on three different simulators including IoTSim-Edge \cite{Jha} and PureEdgeSim \cite{Mechalikh} along with iFogSim2. These simulators have been selected because of their recent inclusions in the literature and open source license. Furthermore, their Java programming language-based implementation does not arise any compatibility issues with the proposed iFogSim2 simulator, which also helps in reconstructing the experimental results. A brief discussion of the simulators is given below. 
	\begin{itemize}
		\item \textbf{IoTSim-Edge:} This monolithic simulator supports the imitation of microservices in form of microelements and provides support for customising the user mobility; however, lacks abstractions for node clustering.   
		\item \textbf{PureEdgeSim:} It supports the modularisation of different simulation components and facilities qualitative allocation of tasks using a built-in Fuzzy inference engine; nevertheless, barely provides functionalities for node clustering and microservice management.     
		\item \textbf{iFogSim2:} The proposed simulator is well-equipped with APIs and built-in policies for illustrating mobility, microservice, and node clustering-related use cases in Edge/Fog computing environments. The function of its different components can also be tuned as per the case studies to create variations in the simulations.      
	\end{itemize}
	\subsubsection{Results}
	The simulation experiments on the CDC use case study are exclusively exploited to demonstrate the efficacy of IoTSim-Edge, PureEdgeSim, and iFogSim2 simulators in terms of supporting mobility, microservice, and node clustering issues. The results are discussed below. 
	\par$\bullet$\textbf{ RAM usage:} Table \ref{Tab_ramUsage} illustrates the RAM usage of different simulators for varying simulation configurations. As noted, IoTSim-Edge supports the imitation of device mobility and microservice orchestration in Edge/Fog computing environments. However, as it does not facilitate modularisation of the simulation components and mostly operates in a monolithic manner, its RAM usage does not vary for \textit{Mobility} and \textit{Mobility+Microservices} simulation configurations. Conversely, PureEdgeSim only supports the \textit{Mobility} configuration. Nevertheless, this simulator consumes more RAM than other simulators because of its built-in Fuzzy inference engine, supporting task allocation based on qualitative features. On the other hand, iFogSim2 supports a wide range of simulation configurations, including \textit{Mobility}, \textit{Mobility+Microservices}, \textit{Mobility+Clustering}, \textit{Microservices+Clustering} and \textit{Mobility+Microservices+Clustering}. Despite facilitating such configurations, the RAM usage for iFogSim2 does not increase significantly because of its modular architecture and lightweight built-in service and resource management policies.      
	\begin{table}[!t] 
		\caption{RAM usage for different simulator}\label{Tab_ramUsage} 
		\centering
		\scriptsize
		\begin{tabular}{|p{3.3 cm}p{1.4 cm}p{1.2 cm}p{1.1 cm}|}
			\hline
			Simulators $\Rightarrow$ \newline \newline Variations $\Downarrow$  & IoTSim-Edge & PureEdge-Sim & iFogSim2 \\
			Mobility & 26\% & 38\% & 12\% \\
			Mobility+Microservices & 26\% & - & 21\% \\
			Mobility+Clustering & - & - &  19\% \\ 
			Microservices+Clustering & - & - & 15\%   \\
			Mobility+Microservices+Clustering & - & - & 32\% \\\hline
		\end{tabular} 
	\end{table}
	\par
\begin{figure}[!t]
		\centering 
		\includegraphics[width=\columnwidth]{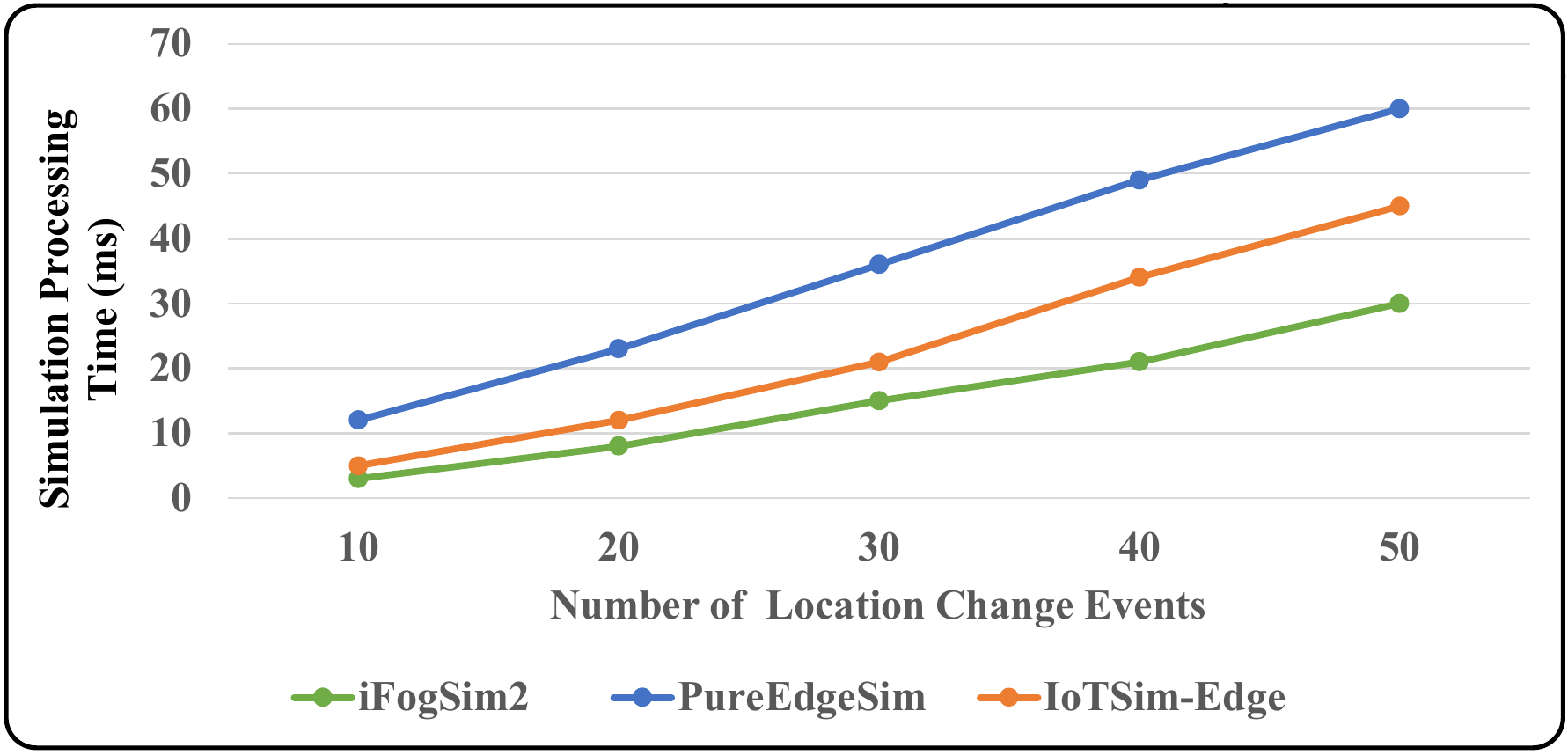}
		\caption{Simulation processing time for different simulators}
		\label{Fig:procMobTime}
	\end{figure}

$\bullet$\textbf{ Simulation time: } Fig. \ref{Fig:procMobTime} depicts the simulation time of IoTSim-Edge, PureEdgeSim and iFogSim2 for \textit{Mobility} configuration. As PureEdgeSim deliberately exploits the Fuzzy inference for managing mobility, it requires more time to imitate the effect of changing locations, which also elevates with the increasing number of such events. However, iFogSim2 performs well in this case because of its low complexity built-in mobility management techniques. Its modular architecture further helps to outperform IoTSim-Edge, which cannot ensure mobility management in a segmental manner.               

	\section{Conclusions and Future Work} \label{sec_conclusion}
	Efficient resource management in Edge/Fog computing environment is an important challenge due to the dynamic and heterogeneous nature of Edge/Fog nodes and IoT devices. In this paper, we put forward iFogSim2 simulator, which is an extension of the iFogsim simulator, to address service migration for different mobility models of IoT devices, distributed cluster formation among Edge/Fog nodes of different hierarchical tiers, and microservice orchestration. To support different simulation scenarios, the new components of the iFogSim2 simulators are loosely coupled, so that components (Mobility, Clustering, and Microservices) can be solely used for the simulation, or they can be integrated for more complex scenarios. Besides, to enhance the usability of iFogSim2, several case studies and test scripts are implemented and integrated with this simulator, which simplifies the process of defining new policies and case studies for its users. The results demonstrate the effectiveness of using iFogSim2 for different case studies and also prove its low footprint compared to other related simulators.
	\par
	As a future work, iFogSim2 simulator can be further improved by integration of monetary-based policies, simulating distributed ledgers across Fog nodes, setting different communication profiles for sensors such as LoRa and Bluetooth, and simulating distributed and federated machine learning approaches.
	
	
	%

	

	\section*{Software Availability}
	The source code of the iFogSim2 simulator is accessible from:
	\href{https://github.com/Cloudslab/iFogSim}{https://github.com/Cloudslab/iFogSim}

	\bibliographystyle{IEEEtran}
	
	\bibliography{reference}


\end{document}